\documentclass[showkeys,superscriptaddress,preprintnumbers,amsmath,amssymb,showpacs,twocolumn,pre]{revtex4-1}
\usepackage[utf8]{inputenc}

\usepackage{graphicx}
\graphicspath{{images/}}
\usepackage{dcolumn}
\usepackage{bm}
\usepackage{array}
\usepackage{amsmath}
\usepackage{amsfonts}
\usepackage{amssymb}
\usepackage{mathtools}
\usepackage{wasysym}
\usepackage{hyperref}
\allowdisplaybreaks

\def\equationautorefname~#1\null{%
  Eq.~(#1)\null
}
\usepackage{color}
\definecolor{Red}{rgb}{1.00, 0.00, 0.00}

\newcommand{\M}{\mathrm{\textbf{M}}}
\newcommand{\x}{\hat{\mathrm{\textbf{x}}}}
\newcommand{\y}{\hat{\mathrm{\textbf{y}}}}
\newcommand{\F}{\mathrm{\textbf{F}}}
\newcommand{\OO}{\mathrm{\textbf{O}}}
\newcommand{\A}{\mathrm{\textbf{A}}}
\newcommand{\B}{\mathrm{\textbf{B}}}

\newcommand{\U}{\mathrm{\textbf{U}}}
\newcommand{\xx}{\mathrm{\textbf{x}}}
\newcommand{\K}{\mathrm{\textbf{K}}}
\newcommand{\MS}{\mathrm{\textbf{S}}}
\newcommand{\Mg}{\mathrm{\textbf{g}}}
\newcommand{\dd}{\mathrm{\textbf{d}}}
\newcommand{\bb}{\mathrm{\textbf{b}}}
\newcommand{\rr}{\mathrm{\textbf{r}_{0}}}

\newcommand{\etheta}{\hat{\mathrm{\textbf{e}}}_\theta}
\newcommand{\stress}{\mathbf{\mathrm{\sigma}}}
\newcommand{\strain}{\mathbf{\mathrm{\epsilon}}}

\usepackage{stackengine}

\usepackage{siunitx}
\begin{document}

\title{Effect of architecture disorder on the elastic response of two-dimensional lattice materials}

\author{Antoine Montiel}
\affiliation{Université Paris-Saclay, CEA, CNRS, SPEC, 91191, Gif-sur-Yvette, France}
\author{Thuy Nguyen}
\affiliation{Léonard de Vinci Pôle Universitaire, Research Center, 92916 Paris La Défense, France}
\affiliation{Université Paris-Saclay, CEA, CNRS, SPEC, 91191, Gif-sur-Yvette, France}
\author{Cindy L. Rountree}
\affiliation{Université Paris-Saclay, CEA, CNRS, SPEC, 91191, Gif-sur-Yvette, France}
\author{Valérie Geertsen}
\affiliation{Université Paris-Saclay, CEA, CNRS, NIMBE, 91191, Gif-sur-Yvette, France}
\author{Patrick Guenoun}
\affiliation{Université Paris-Saclay, CEA, CNRS, NIMBE, 91191, Gif-sur-Yvette, France}
\author{Daniel Bonamy}
\affiliation{Université Paris-Saclay, CEA, CNRS, SPEC, 91191, Gif-sur-Yvette, France}


\begin{abstract}
We examine how disordering joint position influences the linear elastic behavior of lattice materials via numerical simulations in two-dimensional beam networks. Three distinct initial crystalline geometries are selected as representative of mechanically isotropic materials low connectivity, mechanically isotropic materials with high connectivity, and mechanically anisotropic materials with intermediate connectivity. Introducing disorder generates spatial fluctuations in the elasticity tensor at the local (joint) scale. Proper coarse-graining reveals a well-defined continuum-level scale elasticity tensor. Increasing disorder aids in making initially anisotropic materials more isotropic. The disorder impact on the material stiffness depends on the lattice connectivity: Increasing the disorder softens lattices with high connectivity and stiffens those with low connectivity, without modifying the scaling between elastic modulus and density (linear scaling for high connectivity and cubic scaling for low connectivity). Introducing disorder in lattices with intermediate fixed connectivity reveals both scaling: the linear scaling occurs for low density, the cubic one at high density, and the crossover density increases with disorder. Contrary to classical formulations, this work demonstrates that connectivity is not the sole parameter governing elastic modulus scaling. It offers a promising route to access novel mechanical properties in lattice materials via disordering the architectures.      
\end{abstract}

\date{\today}



\maketitle

\section{Introduction}\label{Sec1}

Cellular, reticulated, truss and lattice materials can exhibit remarkable stiffness-to-weight ratio \cite{Gibson1997,Fleck2010,Yu2018,Zhang2019}. Nature exemplifies this phenomenon, e.g. in the cellular structure of wood \cite{Ando1999}, trabecular bones \cite{Ryan2013}, plant parenchyma \cite{Liedekerke2010} and sponges \cite{Shen2013}. Aerogels, metallic foams and bio-inspired lightweight cellular materials also find a broad range of applications in industry, with transportation and aerospace driving the field. Still, the large porosity of these materials inevitably causes substantial reduction in the mechanical properties. The stiffness of a stochastic foam with a relative density of $1\%$ is about a millionth of that of its constituent material \cite{Moner-Girona1999,Ma2000}.   

Recent and formidable progress in additive manufacturing and 3D printing boosted research in the field \cite{Yu2018,Zhang2019}, giving way to the fabrication of high precision micro-/nanoarchitectured cellular materials of complex geometries \cite{Zhang2019}. In this context, micro-/nanolattice materials consisting of periodically arranged beams or tubes of micrometer/nanometer dimensions exhibit unprecedented stiffness-to-weight ratio \cite{Schaedler2011,Meza2014,Meza2015,Han2015,Zheng2016,Meza2017}. In general, the Young's modulus, $E$, typically scales as the density cubed $\rho^3$ in foams, aerogels and other cellular materials with randomly distributed porosity\cite{Ma2001,Worsley2009,Tillotson1992}. On the other hand, $E$ scales as $\rho^2$ in periodic hollow-tube microlattices with octahedral basic cells \cite{Schaedler2011}, as $\rho^{1.6}$ in octet-truss geometry \cite{Meza2014}, or even linearly with $\rho$ for well-chosen hierarchical architectures \cite{Meza2015, Zheng2016}. 

The $E~vs.~\rho$ scaling finds its origin in the deformation mode of the lattice \cite{Deshpande2001,Evans2001,Ashby2005}. When deformation is dominated by the bending of the constituent beams, $E \sim (s/\ell)^4$ ($E \sim (s/\ell)^3$ in 2D) where $s$ and $\ell$ are the typical cross-sectional size and length of solid beams. When the lattice deformation arises from beam stretching or compression, $E \sim (s/\ell)^2$ ($E \sim s/\ell$ in 2D). Relative density goes as $\rho \sim (s/\ell)^2$ ($\rho \sim s/\ell$ in 2D). Hence, $E \sim \rho^2$ ($E\sim \rho^3$ in 2D) in bending-dominated lattices, and $E \sim \rho$ (likewise in 2D) in stretching-dominated lattices. To assess whether a lattice material is stretching- or bending-dominated one has to consider the collapse mechanisms in a pin-jointed structure of the same geometry \cite{Deshpande2001, Ongaro2018}. Periodic pin-jointed structures of node connectivity $Z < 6$ ($Z <4$ in 2D) do not satisfy Maxwell's conditions and are non-rigid \cite{Maxwell1864,Deshpande2001}. Consequently, the deformations in the parent welded-joint lattice material  are governed by the beam rotation at the nodes, and bending dominates elastic behavior. Pin-jointed frames with $Z \geq 12$ ($Z \geq 6$ in 2D) possess no collapse mechanism, in the sense that any deformation generates an increase of the strain energy \cite{Pellegrino1986,Deshpande2001}. They are fully-rigid, the associated lattice materials are predicted to be stretching-dominated, and the Gurtner-Durand bound provides a maximum achievable Young's modulus in isotropic structures \cite{Gurtner2014}. Note that this predicts $E \sim \rho$ in Octet-truss lattices ($Z=12$). This is in apparent contradiction with the experimental observation of $E \sim \rho^{1.6}$ reported in Ref. \cite{Meza2014}. This discrepancy is discussed in Ref. \cite{Meza2014} and attributed to the hollowness of the tubes, affecting the structural integrity of the nodes and yielding an effective connectivity smaller than 12. 

Finally, periodic structures of intermediate connectivity $6\leq Z<12$ ($4\leq Z<6$ in 2D) are referred to as periodically rigid. The pin-jointed version of these lattices obeys Maxwell conditions, but there exists at least one periodic mechanism causing them to collapse. In the Kagome structure \cite{Kapko2009}, this collapse mechanism does not produce macroscopic strain \cite{Ongaro2018} and the parent welded-joint lattice exhibits a stretching-dominated behavior when loaded in any direction. Conversely, the square lattice is bending-dominated when loaded in the diagonal direction. This is because its pin-jointed version collapses upon such loading. 

Literature exemplifies a number of works concerning the mechanics of lattice materials with periodic (crystalline) geometries. However, only a limited number of studies examined disordered or non-periodic architectures \cite{OstojaStarzewski1989, Silva1995, Silva1997, Symons2008, Mukhopadhyay2017, Pham2019, RayneauKirkhope2019}. Hence, the present study focuses on lattices with constant node connectivity, with a large range of beam aspect ratios and levels of geometrical disorder. Starting from 2D crystalline lattices, increasing levels of disorder are introduced. Subsequently, studies herein examine how this modifies their elasticity response via numerical simulations on beam networks. Sec. \ref{Sec2} presents the method. In this regard, different initial geometries are selected as representative of mechanically isotropic/anisotropic structures, and low, intermediate and large connectivity. Section \ref{Sec3}A  analyzes the impact of disorder on the spatial distribution of stress and strain fields at the local scale, and subsequently Sec. \ref{Sec3}B analyzes its impact on the local compliance tensor. Special emphasize is paid to determine the correlation length associated with the spatial fluctuations, and subsequently the continuum-level scale elasticity tensor. Section \ref{Sec3}C concentrates on how anisotropy evolves with increasing disorder. Lastly, Sec. \ref{Sec3}D looks at how disorder affects the elastic response over a broad range of beam aspect ratios. Additionally, it investigates how disorder affects the Young's modulus prefactor and Poisson's ratio in isotropic lattices. Section \ref{Sec4} discusses the results, and Sec. \ref{Sec5} provides a brief conclusion.


\section{Simulation framework}\label{Sec2}

\subsection{Lattice geometry}

All lattice specimens studied hereafter are enclosed within disks of radius $R$. Starting meshes are summarized in Tab. \ref{table:geometry}. Nodes are connected by elements of length $\ell$ and cross-section size $s$. Clamping conditions are prescribed at the nodes, and there is an energy cost associated with node rotation. 

\begin{table}[tbp]
\centering
\begin{tabular}{| c | c | c | c |} 
\hline 
Mesh type & Node connectivity & Isotropic & Deformation mode \\
\hline
$\hexagon$ & 3 & Yes & Bending \\
$\square$ & 4 & No & Mixed \\
$\triangle$ & 6 & Yes & Stretching \\
\hline
\end{tabular}
\caption{Deformation mode and isotropy in the 2D crystalline lattices studied here, in absence of introduced disorder (from \cite{Gibson1997}).}
\label{table:geometry} 
\end{table}

\begin{figure}[tbp]
\centering 
\includegraphics[width=\columnwidth]{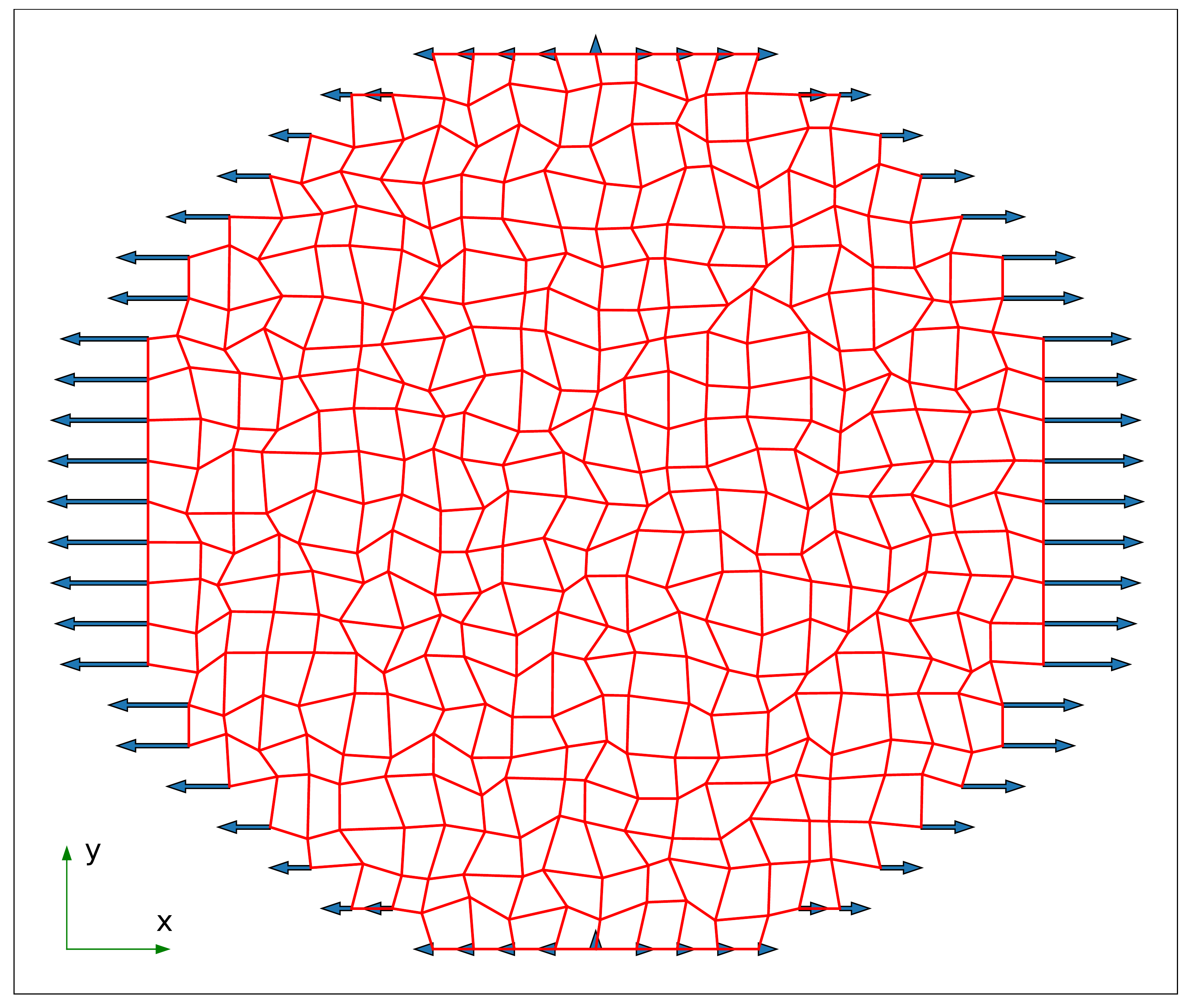}
\caption{(color online) Square lattice disordered by displacing randomly every node by a prescribe distance $u=0.3 \ell_0$ along a randomly chosen direction. Tensile loading (Thick horizontal blue arrows) is applied by imposing a prescribed displacement $\U(\M_b)$ at every node $\M_b$ on the lattice boundary (cosine dependency, see Eq. \ref{eq:loading}).}
\label{fig:Lattice_des}
\end{figure}

To  gradually move from an ordered (crystalline) to a disordered (amorphous) structure, we proceeded as follow: Nodes are first placed in a 2D periodic configuration. This sets the initial crystalline lattice. Then, a tunable disorder is introduced by displacing every node by a prescribed distance $u$ along a randomly chosen direction. Figure \ref{fig:Lattice_des} shows a typical snapshot of a specimen obtained following this procedure. The disorder intensity is set by $u$, which lies in the range $0 \leq u < 0.5\ell_0$, so that struts cannot overlap. Three initial periodic lattices were selected: triangular lattice that is mechanically isotropic and fully rigid (connectivity $Z=6$), honeycomb lattice that is isotropic and non-rigid ($Z=3$), and square lattice that is anisotropic and periodically rigid ($Z=4$).     

Henceforth, the lattice architecture is given by four control parameters: geometry (triangular, honeycomb or square), element length $\ell_0$ and element aspect ratio $s/\ell_0$ in the parent crystalline lattice, and level of disorder $u$. The specimen size is set by the specimen radius $R$. In the following, $\ell_0$ is chosen equal to unity and $R > 30 \ell_0$.

\subsection{Mechanical test simulation}

In the following, $\{\x,\y\}$ refers to the global frame. $\{\x_l,\y_l\}$ refers to the local frame at the considered element, so that $\x_l$ and $\y_l$ are respectively parallel and perpendicular to this element. A beam model was used to determine the lattice deformation in response to a prescribed loading. Each beam has a length $\ell$, a square-section of side length $s$ and are made of an isotropic bulk material of Young's modulus $E_s$ and Poisson's ratio $\nu_{s}$. The state of each node is given by three parameters: the two components of the displacement vector $\U$ and the rotation $\phi$. 

Timoshenko-Ehrenfest beam theory \cite{Goda2012} is used to relate the axial force $N$, shear force $V$, and torque $M$ applied on beam elements to the nodal displacements $\U$ and rotations $\phi$. The local stiffness matrix is:

\begin{equation}
\small
\K_l
=
\begin{bmatrix} 
\frac{E_{s}S}{\ell} & 0 & 0 & \frac{-E_{s}S}{\ell} & 0 & 0 \\
0 & \frac{12E_{s}I}{(1+\lambda)\ell^3} & \frac{6E_{s}I}{(1+\lambda)\ell^2} & 0 & \frac{-12E_{s}I}{(1+\lambda)\ell^3} & \frac{6E_{s}I}{(1+\lambda)\ell^2} \\
0 & \frac{6E_{s}I}{(1+\lambda)\ell^2} & \frac{(4+\lambda)E_{s}I}{(1+\lambda)\ell} & 0 & \frac{-6E_{s}I}{(1+\lambda)\ell^2} & \frac{(2-\lambda)E_{s}I}{(1+\lambda)\ell} \\
\frac{-E_{s}S}{\ell} & 0 & 0 & \frac{E_{s}S}{\ell} & 0 & 0 \\
0 & \frac{-12E_{s}I}{(1+\lambda)\ell^3} & \frac{-6E_{s}I}{(1+\lambda)\ell^2} & 0 & \frac{12E_{s}I}{(1+\lambda)\ell^3} & \frac{-6E_{s}I}{(1+\lambda)\ell^2} \\
0 & \frac{6E_{s}I}{(1+\lambda)\ell^2} & \frac{(2-\lambda)E_{s}I}{(1+\lambda)\ell} & 0 & \frac{-6E_{s}I}{(1+\lambda)\ell^2} & \frac{(4+\lambda)E_{s}I}{(1+\lambda)\ell} \\
\end{bmatrix}
\label{eq:stiffness}
\end{equation}

\noindent where $S$ and $I$ are the cross-section area and moment of inertia of the beam element; $\lambda$ is the shear correction factor. For square beams, $S=s^2$, $I=s^4/12$ and $\lambda = (12/5)(1+\nu_{s})(s/\ell)^2$. The term $E_s S/\ell$ in the stiffness matrix is associated with the traction/compression of the element, whereas the terms proportional to $ E_s I/\ell^2$ and $ E_s I/\ell^3$ are associated with shearing and torque. Timoshenko-Ehrenfest beam theory is preferred to Euler-Bernoulli theory since it takes into account the shear deformation of the cross-section and, as such, permits the description of thick beams. The relation between local nodal displacements and rotations $[\dd_l]=[U_{x_l}(\A),\allowbreak U_{y_l}(\A),\allowbreak \phi(\A),\allowbreak U_{x_l}(\B),\allowbreak U_{y_l}(\B), \phi(\B)]^T$ and local forces and moments $[\bb_l]=[F_{x_l}(\A),\allowbreak F_{y_l}(\A),\allowbreak M(\A),\allowbreak F_{x_l}(\B),\allowbreak F_{y_l}(\B),\allowbreak M(\B)]^T$ is:

\begin{equation}
        [\K_l] [\dd_l] = [\bb_l].
\end{equation}

\noindent In this setting, the subscript $l$ refers to local quantities, the superscript $T$ represents the transpose of a vector or matrix, and $\A$ and $\B$ refer to the edge nodes of the considered element. To construct the system of equations for the complete lattice, one needs (1) write each element stiffness matrix in the global coordinate frame (by multiplying it by the appropriate rotation matrix), and (2) subsequently, add the element matrix in the global stiffness matrix as classically done in finite element (FE) or beam models \cite{Nukala2010}. The set of equations describing equilibrium at each node is:

\begin{equation}
        [\K] [\dd] = [\bb],
        \label{eq:problem}
\end{equation}

\noindent Here, $[\bb]=[F_x(\A_1),\allowbreak F_y(\A_1),\allowbreak M(\A_1),\allowbreak F_x(\A_2),\allowbreak F_y(\A_2),\allowbreak M(\A_2),...]^T$ is the load vector, $[\dd] = [U_{x}(\A_1),\allowbreak U_{y}(\A_1),\allowbreak \phi(\A_1),\allowbreak U_{x}(\A_2),\allowbreak U_{y}(\A_2),\allowbreak \phi(\A_2),...]^T$ is the global displacement vector, and $[\K]$ is the global stiffness matrix.

Loading is applied by imposing a displacement on the boundary nodes [Fig. \ref{fig:Lattice_des}]. A loading direction $\theta_{load}$ with respect to $x$  is prescribed. Imposed displacement is then set to unity along this direction, and decreases as a cosine law as the considered direction departs from $\theta_{load}$ :

\begin{equation}
\U(\M_b) = \hat{\U}_{tens/shear}\cos(\theta_b-\theta_{load}),
\label{eq:loading}
\end{equation}

\noindent where $(r_b,\theta_b)$ are the polar coordinates of the considered node $\M_b$ along the specimen boundary. $\hat{\U}_{tens}$ and $\hat{\U}_{shear}$ are unit vectors parallel  and perpendicular to $\theta_{load}$, and are associated with tension and shear loading, respectively. This cosine variation allows a minimization of the impact of boundary discreetness on the stress and strain fields in the bulk lattice. 

The loading conditions above are implemented in Eq. \ref{eq:problem}, by setting $\F(\M_b)=\U(\M_b)$ in  $\{\bb\}$ at the appropriate places and replacing the corresponding blocks in the stiffness matrix $[\K]$ by identity matrix blocks. The final set of equations is solved by inverting the stiffness matrix, using sparse Cholesky decomposition to speed up the process. This provides the displacements $\U(\M)$ and rotation $\phi(\M)$ of each node $\M$ in the loaded lattice.

\subsection{Local stress and strain computation}

The next step determines the continuum stress and strain fields in the sample. For this, we draw inspiration from methods developed to study granular flows \cite{Addetta2002, Bonamy2009}. Voronoi tessellation associates a continuum elementary volume to each node $\M$ of the lattice. At each location $(x,y)$ within the Voronoï polyhedron $P_{vor}(\M)$ centered on $\M$, the local stress tensor  $\mathbf{\mathrm{\sigma}}$ is: 

\begin{equation}
\mathbf{\mathrm{\sigma}} = \frac{1}{2S_{vor}s}\sum_{p} \F_{\M_p\rightarrow\M} \otimes \left(\xx(\M) - \xx(\M_p)\right),
\label{eq:stress}
\end{equation} 

\noindent where $S_{vor}$ is the area of the Voronoï polyhedron, $\M_p$ are the nodes connected to $\M$,  $\xx(\M_p)$ and $\xx(\M)$ are the position of $\M$ and $\M_p$, $\F_{\M_p\rightarrow\M}$ is the force applied by the beam connecting $\M$ and $\M_p$ to $\M$, and $\otimes$ is the vector dyadic product. 

A best-fit algorithm then determines the local strain tensor on the same Voronoi polyhedron. The components $e_{ij}=\partial U_i /\partial x_j$ of the displacement gradient tensor are prescribed so that they minimize:

\begin{equation}
\chi_i = \sum_{p} \left( e_{ij}(\M) \left(x_{j}(\M_p)-x_{j}(\M)\right) - \left(U_i(\M_p)-U_i(\M)\right) \right)^2,
\label{eq:strain1}
\end{equation} 

\noindent where indices $\{i,j\} \in \{x,y\}$. Einstein summation convention is used here on repeated indices. The components of the strain tensor are: 

\begin{equation}
\epsilon_{ij}(\M) = \frac{1}{2}\left( e_{ij}(\M) + e_{ji}(\M)\right) 
\label{eq:strain2}
\end{equation}

\section{Results}\label{Sec3} 

\subsection{Spatial distribution of local stress and strain tensors: influence of disorder}

\begin{figure}[tpb]
\centering 
\includegraphics[width=\columnwidth]{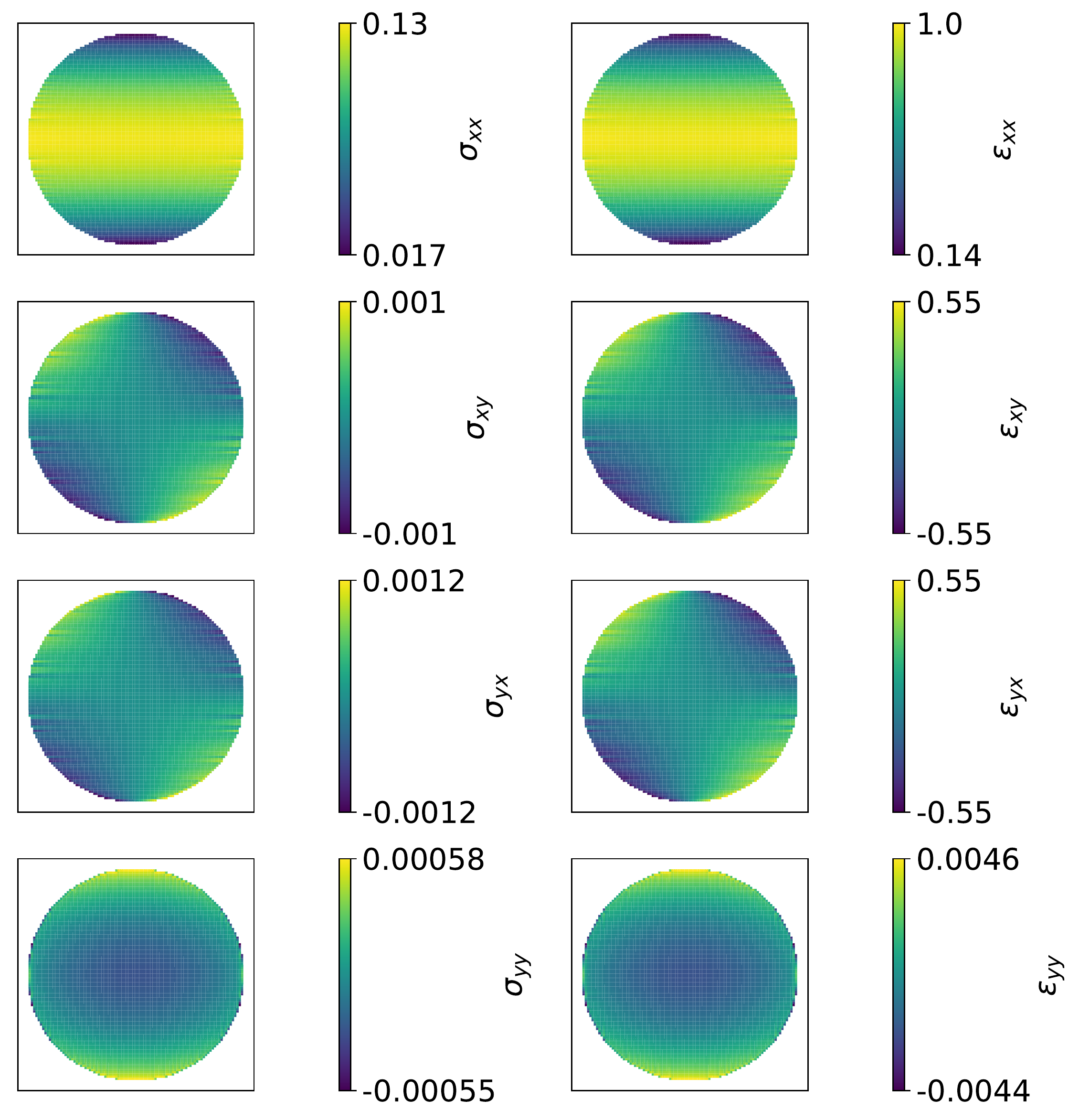}
\caption{(color online) Maps of local stress and strain in a square lattice in absence of architecture disorder. The eight panels correspond to the $\sigma_{xx}$, $\sigma_{xy}$, $\sigma_{yx}$ and $\sigma_{yy}$ (left column), and $\epsilon_{xx}$, $\epsilon_{xy}$, $\epsilon_{yx}$ and $\epsilon_{yy}$ (right column). $\epsilon_{ij}$ are expressed in $\ell_0/R$ units and $\sigma_{ij}$ are expressed in $E_s \ell_0/R$ units. Horizontal tensile loading is imposed by prescribing the displacement at the boundary nodes according to Eq. \ref{eq:loading} with $\theta_{load}=0$. In this simulation, $s/\ell_0=1/8$ and the specimen size is $R=50\ell_0$.}
\label{fig:Map_stress_strain}
\end{figure}
 
Figure \ref{fig:Map_stress_strain} shows local stress and strain maps for a periodic square lattice loaded in tension along $x$ axis. Note that $\sigma_{xy}(x,y) = \sigma_{yx}(x,y)$ everywhere, as expected for a Cauchy stress tensor. This is always observed, regardless of the lattice geometry and amount of disorder introduced. Note also that  $\sigma_{xx}$ is two orders of magnitude larger than $\sigma_{xy}$ and $\sigma_{yy}$. This is due to the fact that Poisson's ratio $\nu_{xy}$ in a square lattice is equal to zero \cite{Gibson1997}. Here we use the notation $\nu_{xy}$ rather than $\nu$ to emphasize that square lattices do not exhibit isotropic elasticity response, but orthotropic one (see next section). Note finally that $\sigma_{xx}(x,y)$ is proportional to $\epsilon_{xx}(x,y)$ everywhere. The prefactor gives the Young modulus of the lattice measured along $x$. It is equal to $E_{x} = E_s s/\ell_0=1/8$, as expected for a square lattice \cite{Gibson1997}.

\begin{figure}[tpb]
\centering 
\includegraphics[width=\columnwidth]{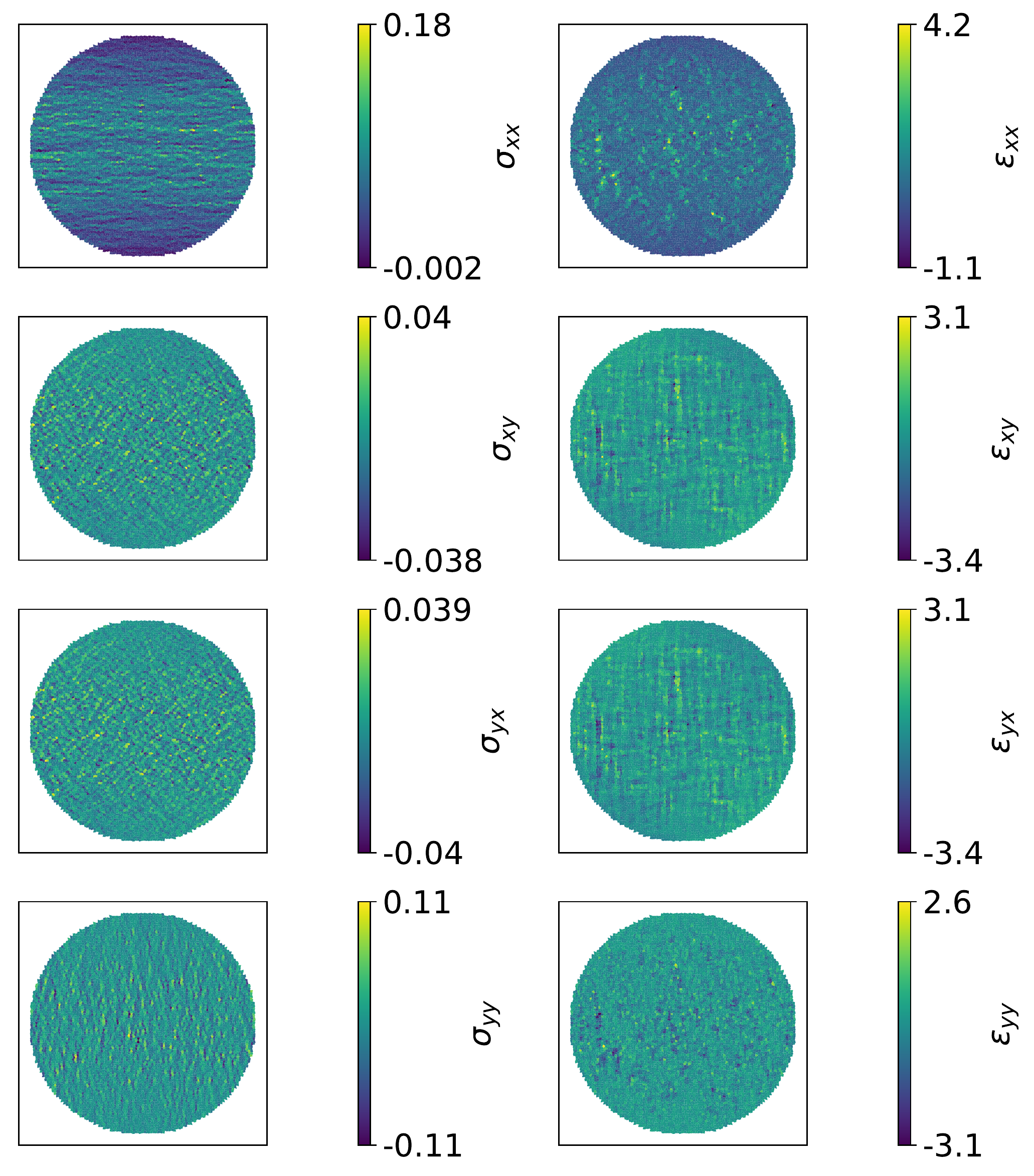}
\caption{(color online) Maps of local stress and strain in a disordered square lattice (disorder intensity: $u=0.3\ell_0$, beam aspect ratio  $s/\ell_0=1/8$, specimen size $R=50\ell_0$). The eight panels correspond to the $\sigma_{xx}$, $\sigma_{xy}$, $\sigma_{yx}$ and $\sigma_{yy}$ (left column), and $\epsilon_{xx}$, $\epsilon_{xy}$, $\epsilon_{yx}$ and $\epsilon_{yy}$ (right column). $\epsilon_{ij}$ are expressed in $\ell_0/R$ units and $\sigma_{ij}$ are expressed in $E_s \ell_0/R$ units. Horizontal tensile loading is imposed by prescribing the displacement at the boundary nodes according to Eq. \ref{eq:loading} with $\theta_{load}=0$.}
\label{fig:Map_stress_strain_des}
\end{figure}

\begin{figure}[tpb]
\centering 
\includegraphics[width=\columnwidth]{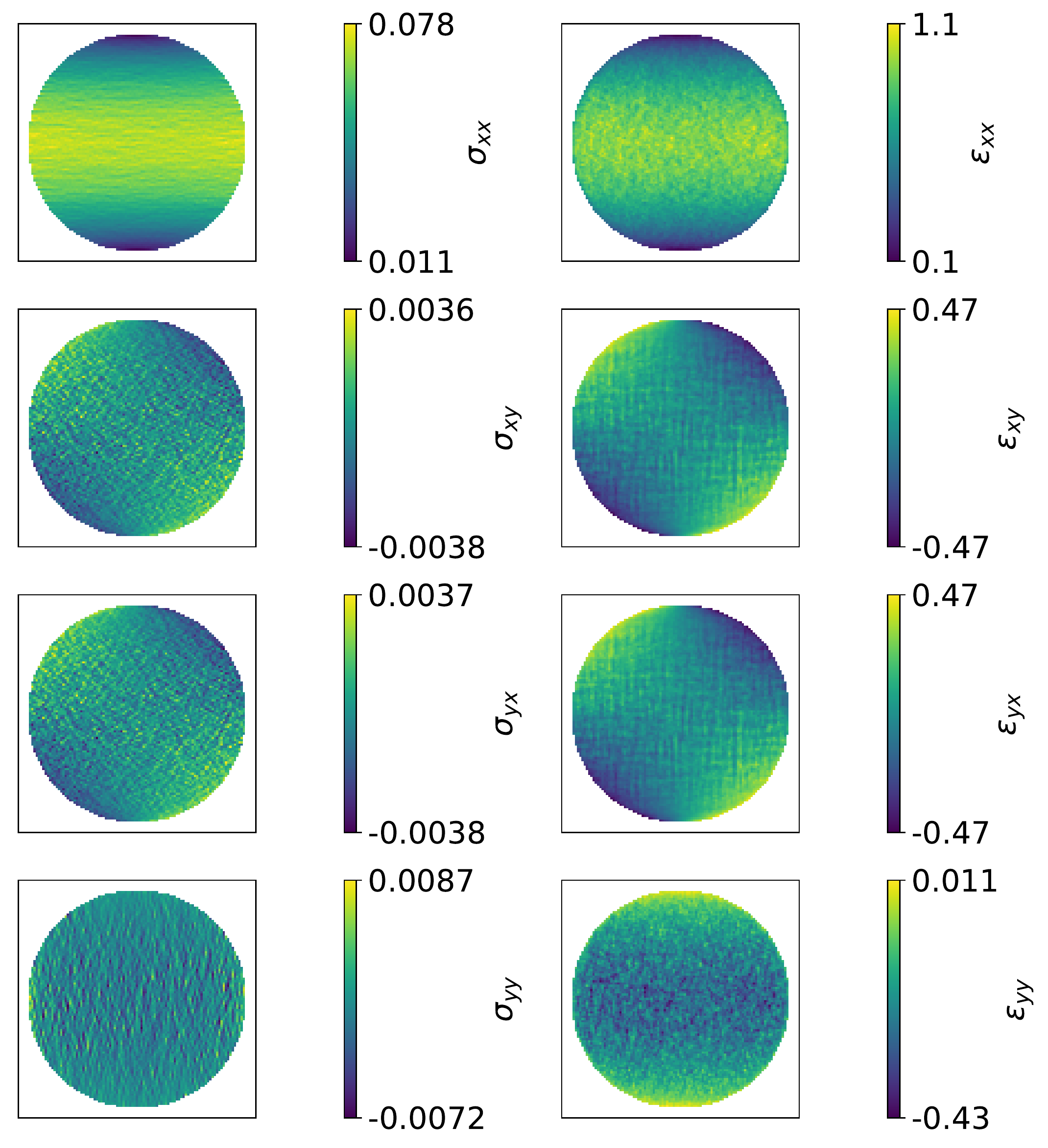}
\caption{(color online) Ensemble averaged maps of local stress (left column) and strain (right column) in a disordered square lattice (disorder amount: $u=0.3\ell_0$, beam aspect ratio  $s/\ell_0=1/8$, specimen size $R=50\ell_0$). The average was taken over $100$ samples with same loading and $u$, but different disorder realizations. The eight panels correspond to the $\sigma_{xx}$, $\sigma_{xy}$, $\sigma_{yx}$ and $\sigma_{yy}$ (left column), and $\epsilon_{xx}$, $\epsilon_{xy}$, $\epsilon_{yx}$ and $\epsilon_{yy}$ (right column). $\epsilon_{ij}$ are expressed in $\ell_0/R$ units and $\sigma_{ij}$ are expressed in $E_s \ell_0/R$ units. Horizontal tensile loading is imposed by prescribing the displacement at the boundary nodes according to Eq. \ref{eq:loading} with $\theta_{load}=0$.}
\label{fig:Map_stress_strain_des_mean}
\end{figure}

Figure \ref{fig:Map_stress_strain_des} presents typical snapshots of the components of local stress and strain tensors in a disordered square lattice. Large spatial inhomogeneities are clearly visible. Note in particular the chain-like structure of the most stressed zones in the top, left panel of Fig. \ref{fig:Map_stress_strain_des}. These resemble force chains observed in granular media \cite{Liu1995}. Chain-like structures are less visible in the strain maps where heterogeneities take the form of relatively isotropic spots [Fig. \ref{fig:Map_stress_strain_des}, top, right panel]. Sec. \ref{Sec3}B takes a closer look at fluctuations to infer relevant length scales which aid in defining a representative elementary volume (REV). Averaging such maps over many configurations of same loading, initial crystal geometry and amount of disorder $u$ (but different realizations) allows smoothening the high frequency fluctuations and reveals the large scale spatial variations of stress and strain fields [Fig. \ref{fig:Map_stress_strain_des_mean}]. The configuration-averaged maps obtained in presence of disorder are significantly different from the maps observed in the pristine crystalline lattice, which show that the introduced disorder changes the elasticity constants.   

\subsection{Spatial distribution of local elasticity constants: on the RVE scale}

The next step is to determine the relevant constants to characterize the elastic response of the lattice at local scale. Using Voigt notation, stress and strain components under plane stress assumption are related via:

\begin{equation}
\begin{bmatrix} \epsilon_{xx} \\\epsilon_{yy} \\2\epsilon_{xy} \\ \end{bmatrix} = \begin{bmatrix} 
S_{11} & S_{12} & S_{14} \\
S_{12} & S_{22} & S_{24} \\
S_{14} & S_{24} & S_{44} \\
\end{bmatrix} \begin{bmatrix} \sigma_{xx} \\\sigma_{yy} \\\sigma_{xy} \\ \end{bmatrix},
\label{eq:Voigt} 
\end{equation} 

\noindent where the symmetric compliance tensor, $\MS$, fully characterizes the material elasticity. For isotropic materials such as triangular and honeycomb lattices, $\MS_{iso}$ is given by:

\begin{equation}
\MS_{iso} = \begin{bmatrix} 
\frac{1}{E} & \frac{-\nu}{E} & 0 \\
\frac{-\nu}{E} & \frac{1}{E} & 0 \\
0 & 0 & \frac{2(1+\nu)}{E}.\\
\end{bmatrix}
\label{eq:Siso} 
\end{equation} 

\noindent The elasticity response is fully characterized by two constants: the Young's modulus $E$ and Poisson's ratio $\nu$. For orthotropic materials such as square lattices, $\MS_{ort}$ is given by:

\begin{equation}
\MS_{ort} = \begin{bmatrix} 
\frac{1}{E_x} & \frac{-\nu_{yx}}{E_y} & 0 \\
\frac{-\nu_{xy}}{E_x} & \frac{1}{E_y} & 0 \\
0 & 0 & \frac{1}{G} \\
\end{bmatrix} \textnormal{, with } \frac{-\nu_{yx}}{E_y} = \frac{-\nu_{xy}}{E_x}
\label{eq:Sort} 
\end{equation} 

\noindent The elasticity response is fully characterized by four constants: Young's moduli $E_x$ and $E_y$ along the $x$ and $y$ axis, Poisson's ratio $\nu_{xy}$, and the shear modulus $G$. Note that in square lattices, $E_x=E_y$ and only three of the sought constants $S_{kl}$ are required.

In the most general situation of disordered architectures, the six constants, $S_{kl}$ of the compliance tensor should be determined. As Eq. \ref{eq:Voigt} only provides three independent equations, a single test is not sufficient to determine them \cite{Tsamados2009}. Hence, for each specimen studied, six different tests were performed: three tensile tests and three shearing tests. For these tests, the imposed external displacements are given by Eq. \ref{eq:loading} and $\theta_{load}=\{0,\pi/4,\pi/2\}$ (square-based lattices) or $\theta_{load}=\{0,\pi/3,\pi/2\}$ (triangular- and honeycomb-based lattices). Each test $p$ provides maps of local stress $\sigma^{(p)}_{ij}(x,y)$ and strain $\epsilon^{(p)}_{ij}(x,y)$. Hence, Eq. \ref{eq:Voigt} provides at each location $(x,y)$ 18 relations between $\epsilon^{(p)}_{ij}$ and $\sigma^{(p)}_{ij}$. These relations involve the six $S_{kl}$ constants needed. The best-fit procedure provides them such that the following equation is minimized for each constant:

\begin{equation}
\begin{split}
\chi = & \sum_{p} \left(\epsilon^{(p)}_{xx} - S_{11}\sigma^{(p)}_{xx} - S_{12}\sigma^{(p)}_{xx} - S_{14}\sigma^{(p)}_{xy} \right)^2\\
           &+ \sum_{p} \left(\epsilon^{(p)}_{yy} - S_{12}\sigma^{(p)}_{xx} - S_{22}\sigma^p_{xx} - S_{24}\sigma^{(p)}_{xy} \right)^2\\
           &+ \sum_{p} \left(2\epsilon^{(p)}_{xy} - S_{14}\sigma^{(p)}_{xx} - S_{24}\sigma^{(p)}_{xx} - S_{44}\sigma^{(p)}_{xy} \right)^2\\
\end{split}
\end{equation} 

\noindent Such homogenization methods are classically used, in FEA, to determine the homogenized elastic constants of complex materials such as composites \cite{Steven1997}.

\begin{figure}[tpb]
\centering 
\includegraphics[width=\columnwidth]{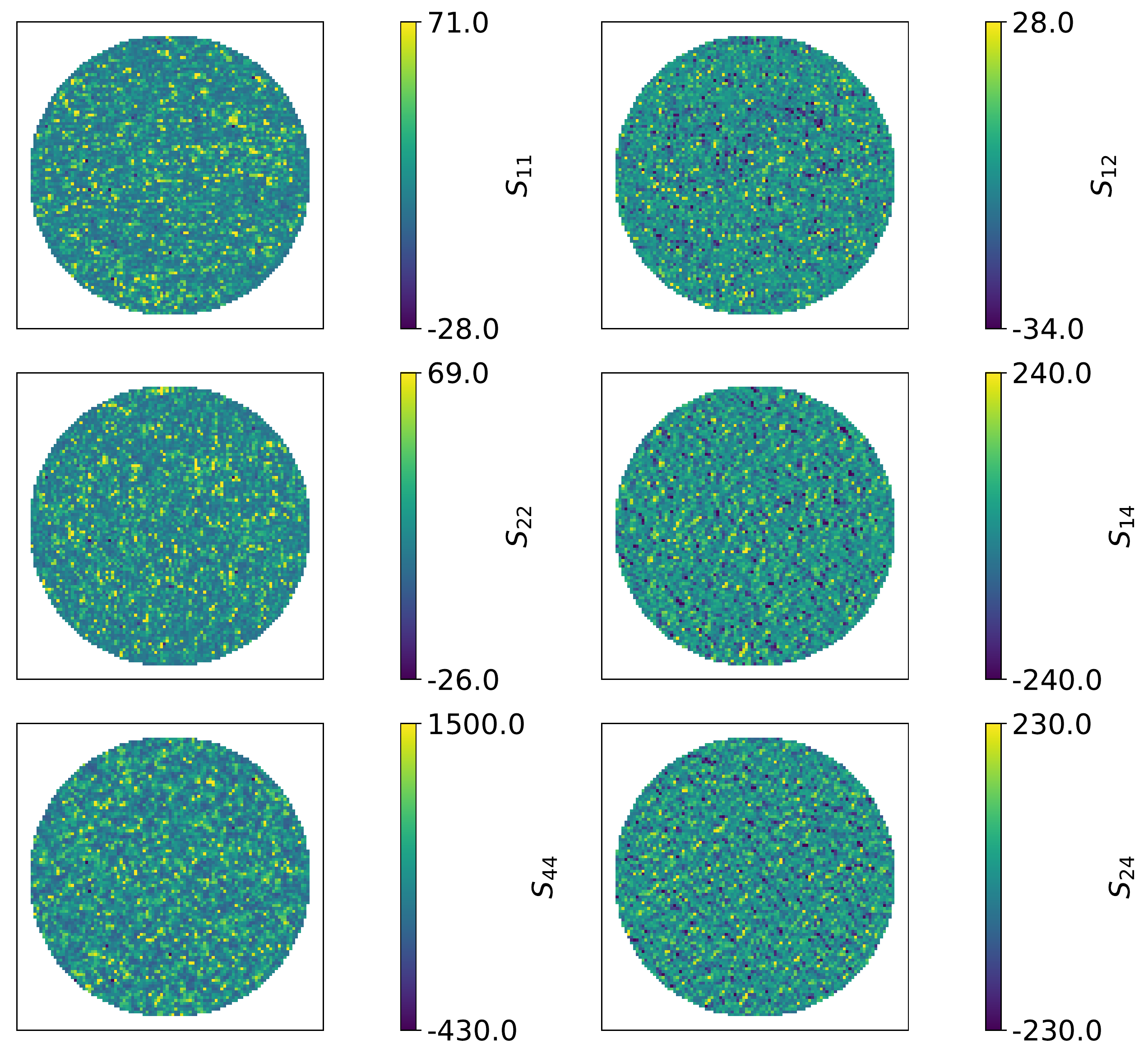}
\caption{(color online) Maps of local compliance tensor $\MS$ in a disordered square lattice (disorder amount: $u=0.3\ell_0$, beam aspect ratio  $s/\ell_0=1/8$, specimen size $R=50\ell_0$). The six panels correspond to the six components $S_{11}$, $S_{12}$, $S_{22}$, $S_{22}$, $S_{24}$, $S_{44}$ [Eq. \ref{eq:Voigt}]. They are all expressed in $E_s$ units.}
\label{fig:S}
\end{figure}

\begin{figure}[tpb]
\centering 
\includegraphics[width=\columnwidth]{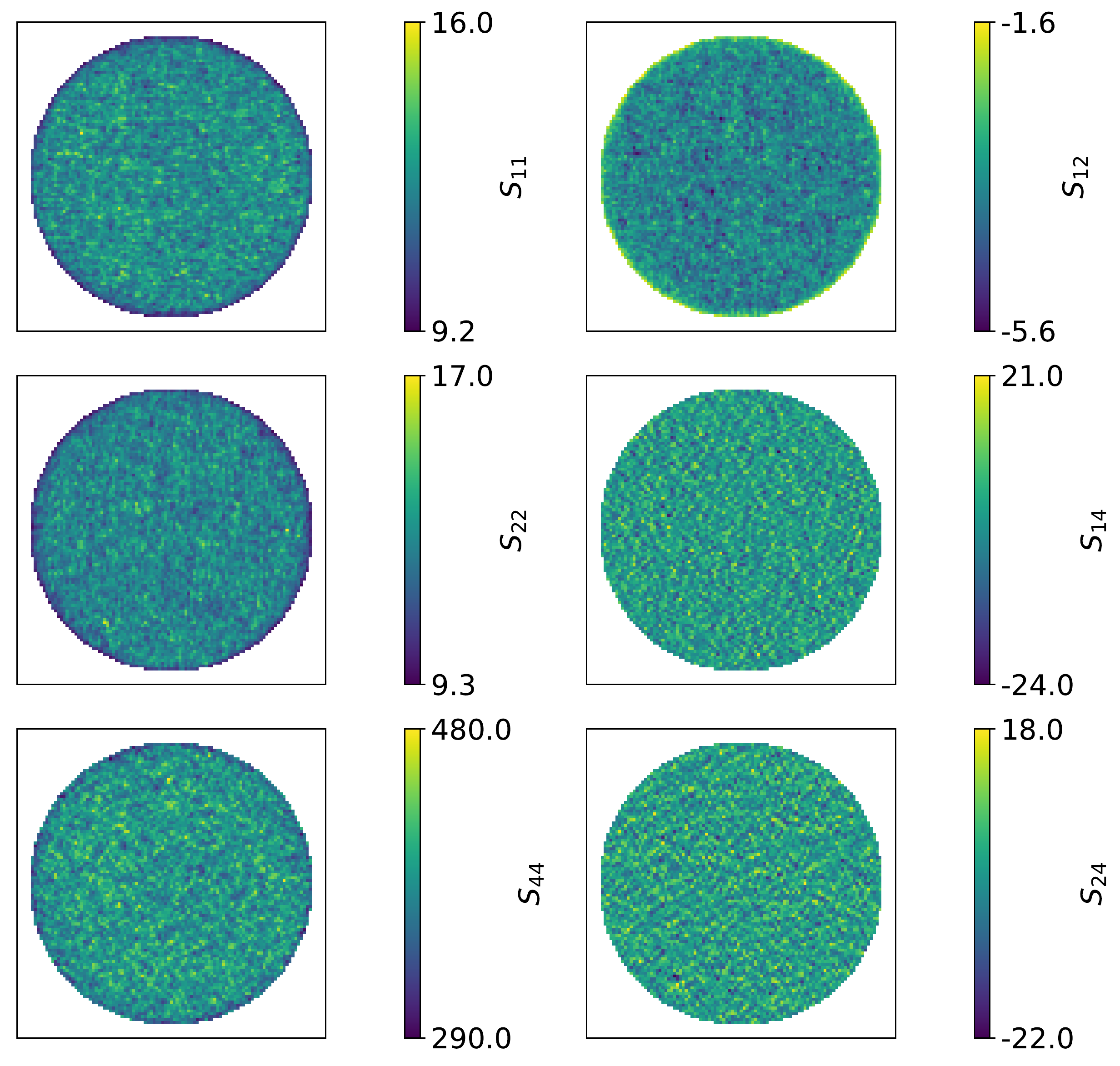}
\caption{(color online)  Ensemble averaged maps of local compliance tensor $\MS$ in a disordered square lattice (disorder amount: $u=0.3\ell_0$, beam aspect ratio  $s/\ell_0=1/8$, specimen size $R=50\ell_0$). The average was taken over $100$ samples of different disorder realisations. The six panels correspond to the six components $S_{11}$, $S_{12}$, $S_{22}$, $S_{22}$, $S_{24}$, $S_{44}$ (Eq. \ref{eq:Voigt}). They are all expressed in $E_s$ units.}
\label{fig:S_mean}
\end{figure}

Figure \ref{fig:S} displays typical maps $S_{kl}(x,y)$ obtained in a disordered square lattice (same lattice material as in Fig. \ref{fig:Map_stress_strain_des}). Note the large spatial variations observed on the $S_{kl}$ maps. Ensemble averaging over 100 samples allows decreasing spatial variability [Fig. \ref{fig:S_mean}]. Still, the maps continue to present the same visual aspects: Except for the edges, these maps are statistically spatially homogeneous, with localized spots uniformly distributed. Note the absence of spatial variations at large wavelength (continuum-level scale), as expected for material constants and contrary to what is observed on the ensemble averaged stress and strain maps [Fig. \ref{fig:Map_stress_strain_des_mean}]

\begin{figure}[tpb]
\centering 
\includegraphics[width=\columnwidth]{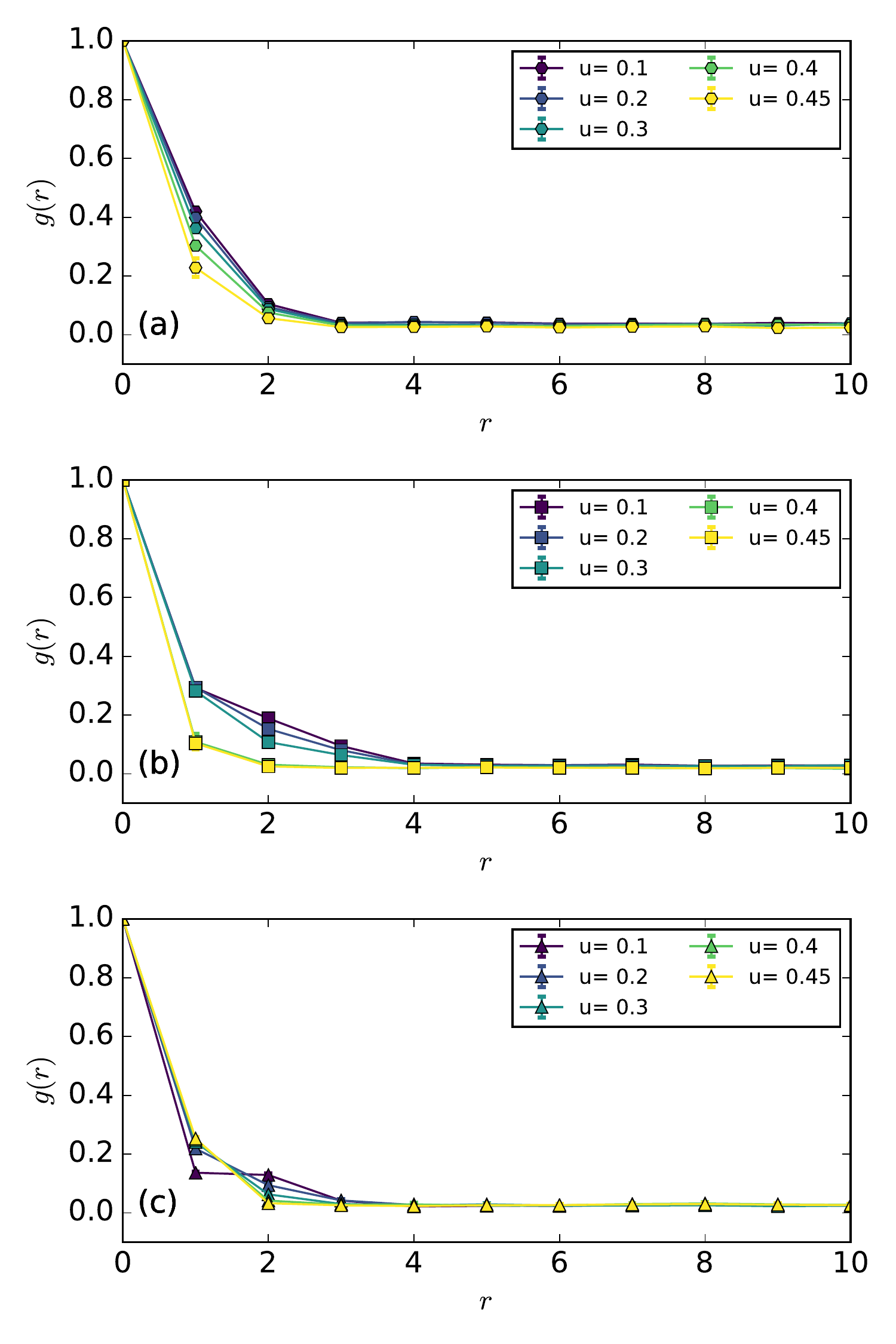}
\caption{(color online) Radial correlation function $g(r)$ at 6 increasing disorder $u$ in honeycomb-based (panel a), square-based (panel b), and triangular-based (panel c). $r$ and $u$ are expressed in $\ell_0$ units. Here, beam aspect ratio is $s/\ell_0=1/8$ and specimen size is $R=40\ell_0$.}
\label{fig:S(R)_vs_R}
\end{figure}

\begin{figure}[tpb]
\centering 
\includegraphics[width=\columnwidth]{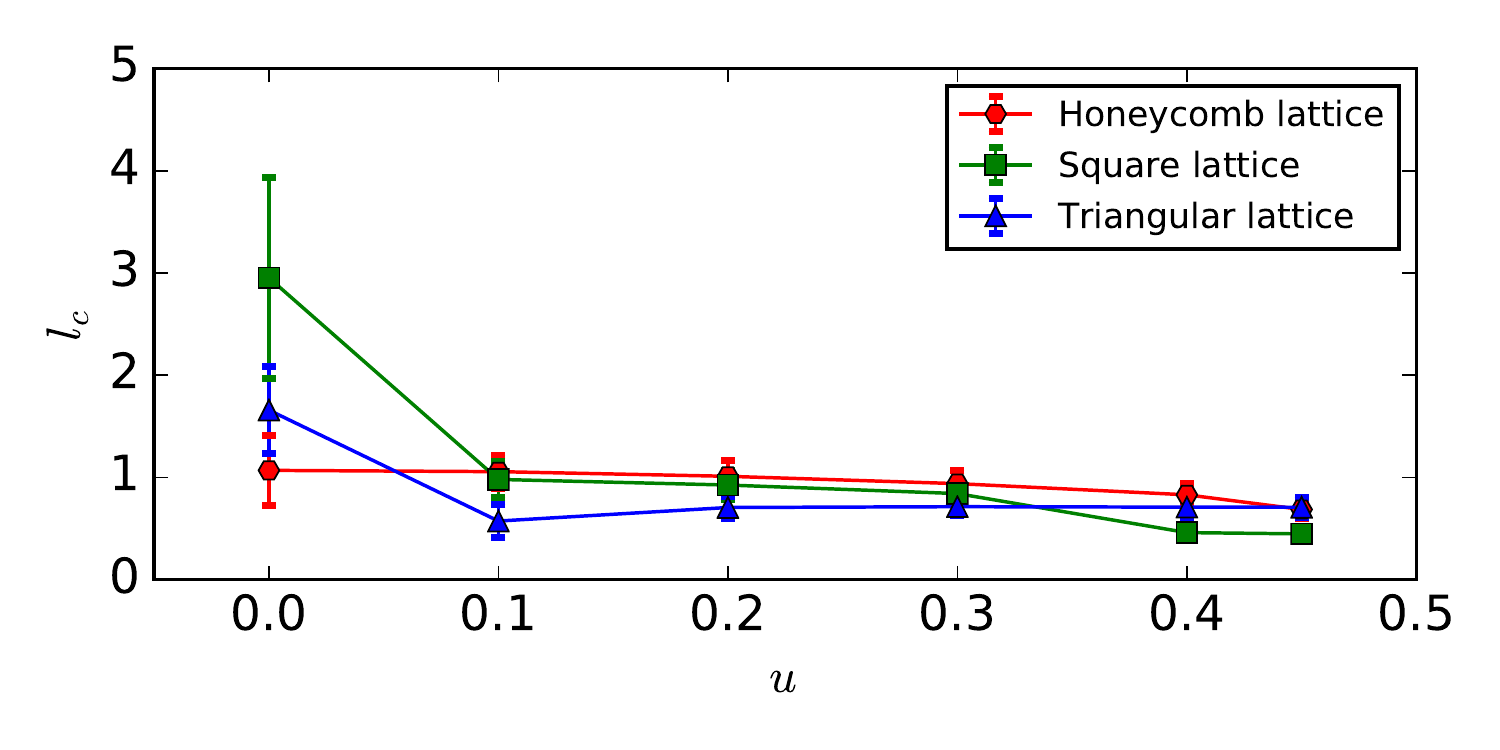}
\caption{(color online) Correlation length $l_c$ as a function of disorder $u$ in the three type of lattice materials. Here, beam aspect ratio is $s/\ell_0=1/8$ and specimen size is $R=40\ell_0$. Both $l_c$ and $u$ are expressed in $\ell_0$ units.}
\label{fig:lc_vs_des}
\end{figure}

To characterize the typical size of the random spots observed in the maps of Fig. \ref{fig:S}, we compute the radial correlation function $g_{kl}(r)$:

\begin{equation}
g_{kl}(r) = \frac{\left\langle \tilde{S}_{kl}(\rr+r \etheta)\tilde{S}_{kl}(\rr)\right\rangle}{\sqrt{\left\langle \tilde{S}^2_{kl}(\rr+r \etheta)\right\rangle\left\langle \tilde{S}^2_{kl}(\rr)\right\rangle}},
\end{equation}

\noindent where $\tilde{S_{kl}}(\rr)=S_{kl}(\rr)-\overline{S}_{kl}$ and $\langle \rangle$ denotes averaging over all positions $\rr$ and subsequently over all direction $\etheta$. $\overline{\MS}$ is the global compliance tensor calculated with averaged stress $\langle \sigma \rangle$ and strain $\langle \epsilon \rangle$. By analogy with $\MS$, $g_{kl}(r)$ forms a symmetric radial correlation matrix $\Mg(r)$. Figure \ref{fig:S(R)_vs_R} presents the variation of its Euclidean norm, $g(r)=||\Mg(r)||$, as a function of $r$ at increasing disorder $u$. Very rapidly, $g(r)$ drops to zero. Fitting these curves by an exponential function $g(r)=\exp(-r/l_c)$ allows defining a correlation length, $l_c$. Its evolution with $u$ is shown in Fig. \ref{fig:lc_vs_des} in the different geometries studied. It is measured to be $\sim \ell_0$ in all cases except the disorder-free square lattice, where $l_c \simeq 3 \ell_0$.
 
A priori, the knowledge of $l_c$ allows setting a REV size $l_{REV}$: Calling $N = l_{REV}/l_c$, one can split the REV into $n=N^d$ elementary cells ($d=2$ in 2D, $d=3$ in 3D) where the quantity of interest is independent and identically distributed (i..i.d.). Calling $\sigma$ the standard deviation of this quantity over one  elementary cell, the central limit theorem tell us that the standard deviation of the same quantity over the REV goes as $\sigma/\sqrt{n}=\sigma/ (l_{REV}/l_c)^{d/2}$ .
As a result, a given quantity coarse-grained over the REV will typically exhibit typical fluctuations that decrease as $l_c/l_{REV}$ in 2D (or $(l_c/l_{REV})^{3/2}$ in 3D). Additional work (see appendix \ref{A1} for details) also shows that when REV size is equal to $\simeq 4 \ell_0$ then global compliance tensor can be calculated directly from macroscopic stresses and strains or alternatively from the averaging of local compliance tensor on REV cells.  As will be discussed in Sec. \ref{sec:locvsglob}, this condition does not imply that the elastic constants are independent of specimen size. In fact, a much larger REV should be prescribed to ensure that these elastic constants are bulk material constants (Sec. \ref{sec:locvsglob} and Fig. \ref{fig:convergence})

The next section analyzes the global compliance tensor $\overline{\MS}$ for different levels of disorder and a large range of aspect ratios.  

\subsection{Anisotropy of elasticity response: influence of disorder}

This section takes a look at how disorder affects the anisotropy of the elasticity response. Initially the global compliance tensor $\overline{\MS}$ is calculated. Next the effective compliance tensor $\overline{\MS}_{eff}$ is defined as:

\begin{equation}
\overline{\MS}_{eff} = \begin{bmatrix} 
\frac{\overline{S}_{11} + \overline{S}_{22}}{2} & \overline{S}_{12} & 0 \\
\overline{S}_{12} & \frac{\overline{S}_{11} + \overline{S}_{22}}{2} & 0 \\
0 & 0 & \frac{\overline{S}_{11} + \overline{S}_{22} + \overline{S}_{44} - 2\overline{S}_{12}}{2} \\
\end{bmatrix}
\end{equation}

\noindent This tensor meets the requirement that $\overline{\MS}=\overline{\MS}_{eff}$ in mechanically isotropic materials. Lastly, the \textit{Universal Anisotropy Index} ($UAI$) is defined by analogy to the Zener index which is limited to the cubic crystals case \cite{Ranganathan2008}:

\begin{equation}
UAI = \frac{||\overline{\MS}_{eff} - \overline{\MS}||}{||\overline{\MS}_{eff}||} 
\label{eq:UAI}
\end{equation} 

\noindent It quantitatively defines how close $\overline{\MS}$ is to $\overline{\MS}_{eff}$. Indeed, with this definition, if the elastic material is isotropic then $UAI=0$, otherwise $UAI>0$. Moreover, as $UAI$ increases, the anisotropy increases. 

\begin{figure}[tpb]
\centering 
\includegraphics[width=\columnwidth]{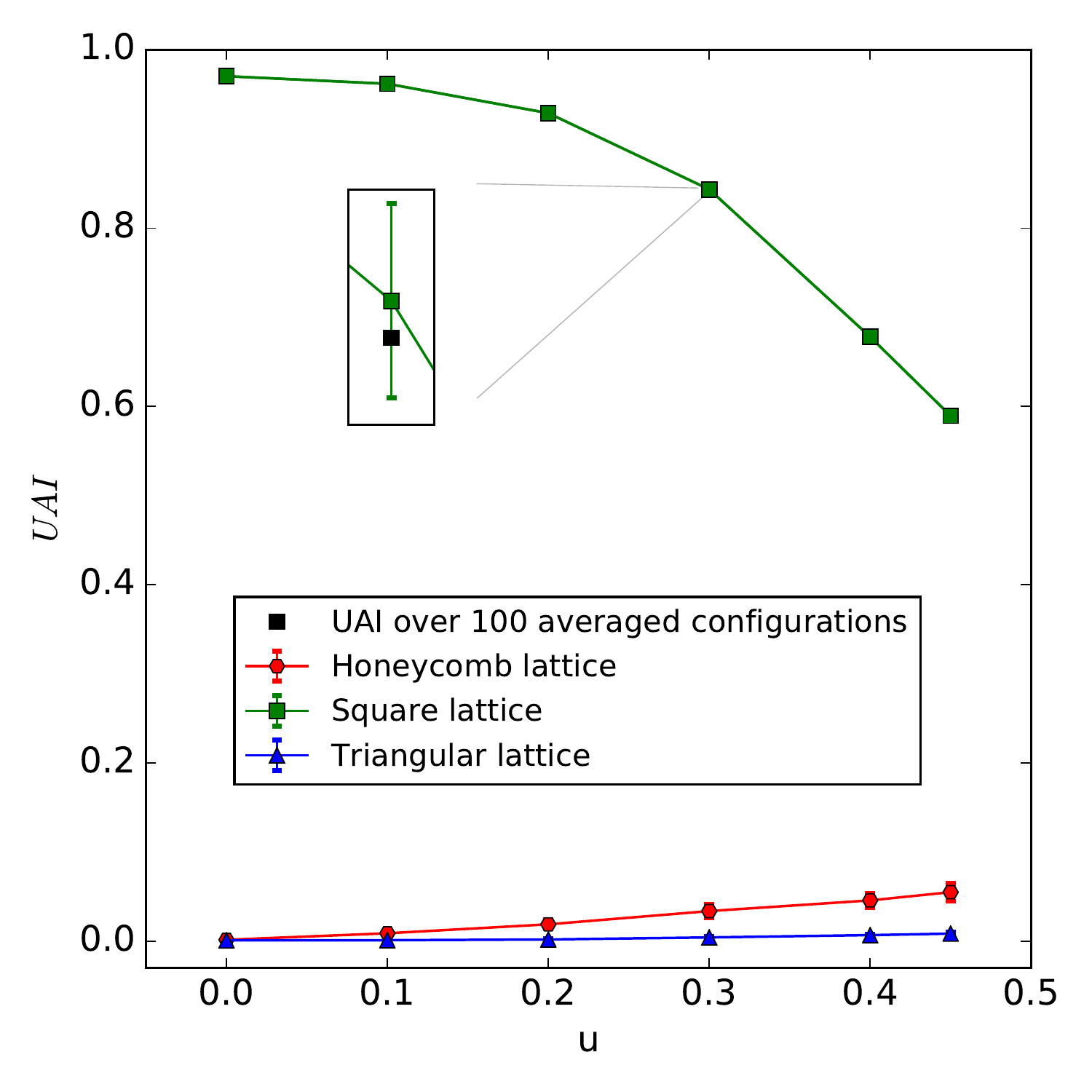}
\caption{Universal anisotropy index $UAI$ (Eq. \ref{eq:UAI}) as a function of disorder level $u$ in the three types of geometries: Honeycomb-based (red $\hexagon$), square-based (green $\square$), and triangular (blue $\triangle$). Here, beam aspect ratio is $s/\ell_0=1/8$ and specimen size is $R=40\ell_0$. $u$ is expressed in $\ell_0$ units. In all the curves, UAI has been obtained from the analysis of a single specimen. Black $\square$ in inset shows the result obtained after having averaged over $100$ configurations in a square-based lattice with $u=0.3$. $u$ is expressed in $\ell_0$ units.}
\label{fig:UAI_vs_disorder}
\end{figure}

Figure \ref{fig:UAI_vs_disorder} presents the evolution of $UAI$ with disorder intensity for the three studied geometries. In absence of disorder, $UAI=0$ in honeycomb and triangular lattice and these lattice materials obey isotropic elasticity, as expected. Increasing disorder has nearly a null effect on the anisotropy index of the already isotropic materials. The largest value in this context is $UAI=5.7\%$, which is observed in the honeycomb lattice at maximum disorder ($u=0.45\ell_0$). Conversely, in absence of architecture disorder, the square lattice is highly anisotropic, with $UAI = 97\%$. In this case, increasing architecture disorder significantly improves the elasticity isotropy, and $UAI$ is nearly half the initial value, $\sim 58\%$, for the maximum disorder intensity.

\subsection{Elasticity versus aspect ratio scaling: effect of disorder on material stiffness}

\begin{table*}[!]
\centering
\begin{tabular}{| l | >{\centering}m{3cm} >{\centering}m{3cm} >{\centering}m{3cm}|} 
\hline 
Lattice geometry & Honeycomb & Square & Triangular \tabularnewline
\hline
Density ($\rho_s$) & $\rho=(2/\sqrt{3})(s/\ell)$ & $\rho=2s/\ell$ & $\rho=2\sqrt{3}(s/\ell)$ \tabularnewline
Young's modulus ($E_s$) & $E=(4/\sqrt{3})(s/\ell)^3$ & $E_x=E_y=s/\ell$  & $E=(2/\sqrt{3})(s/\ell)$  \tabularnewline
Poisson's ratio & $\nu = 1$  & $\nu=0$ & $\nu=1/3$ \tabularnewline
Shear modulus ($E_s$) &  $G=(1/\sqrt{3})(s/\ell)^3$ & $G=(1/2)(s/\ell)^3$ & $G=(\sqrt{3}/4)(s/\ell)$ 
\tabularnewline
\hline
\end{tabular}
\caption{Elasticity constants in 2D crystalline lattices in absence of introduced disorder (from \cite{Gibson1997}).}
\label{table:E_nu_theory} 
\end{table*}

\begin{figure}[tpb]
\centering 
\includegraphics[width=\columnwidth]{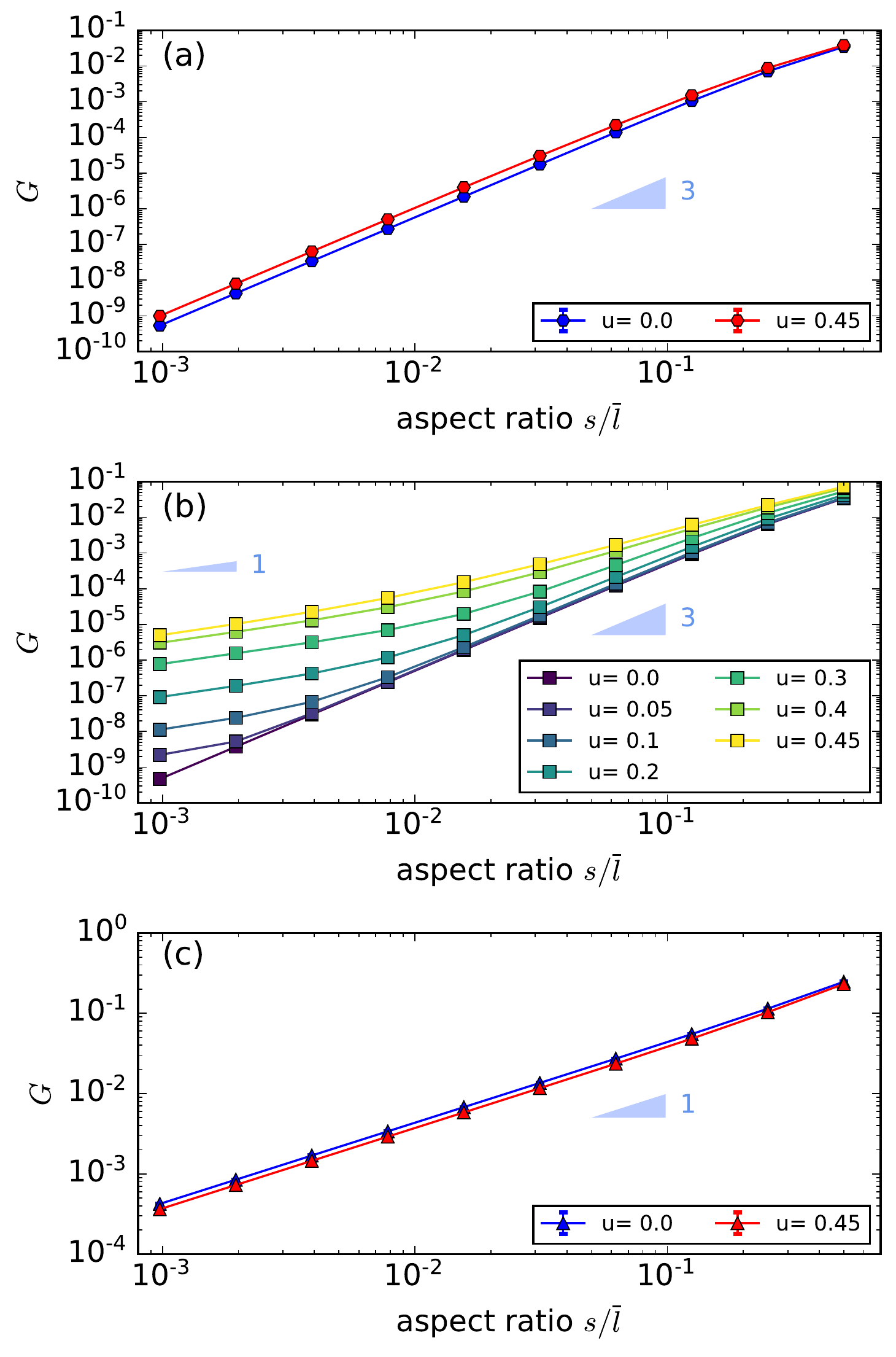}
\caption{(color online) Shear modulus $G$ as a function of beam aspect ratio $s/\ell_0$  with increasing disorder $u$ in honeycomb-based lattices (panel a), square-based lattices (panel b), and triangular-based lattices (panel c). In all panels, axes are logarithmic. In panels (a) and (c), blue (dark gray) curve correspond to $u=0$ (no disorder) and red (light gray) curve corresponds to $u=0.45$ (maximal disorder). In these two panels, the introduction of disorder does not affect the scaling exponent, which remains equal to 3 in honeycomb-based (panel (a)), and to 1 in triangular-based lattice (panel (c)). Conversely, the introduction of disorder affect the scaling in square-based lattice (panel (b)), and the two scaling exponents are observed: $1$ at small aspect ratios and 3 at large ones. The different colors correspond to different levels of disorder $u$ (see caption in panel (b)). Here, specimen size is $R=32\ell_0$. $G$ is expressed in $E_s$ units and $u$ is expressed in $\ell_0$ units.}
\label{fig:S44_vs_ratio}
\end{figure}

\begin{figure}[tbp]
\centering 
\includegraphics[width=\columnwidth]{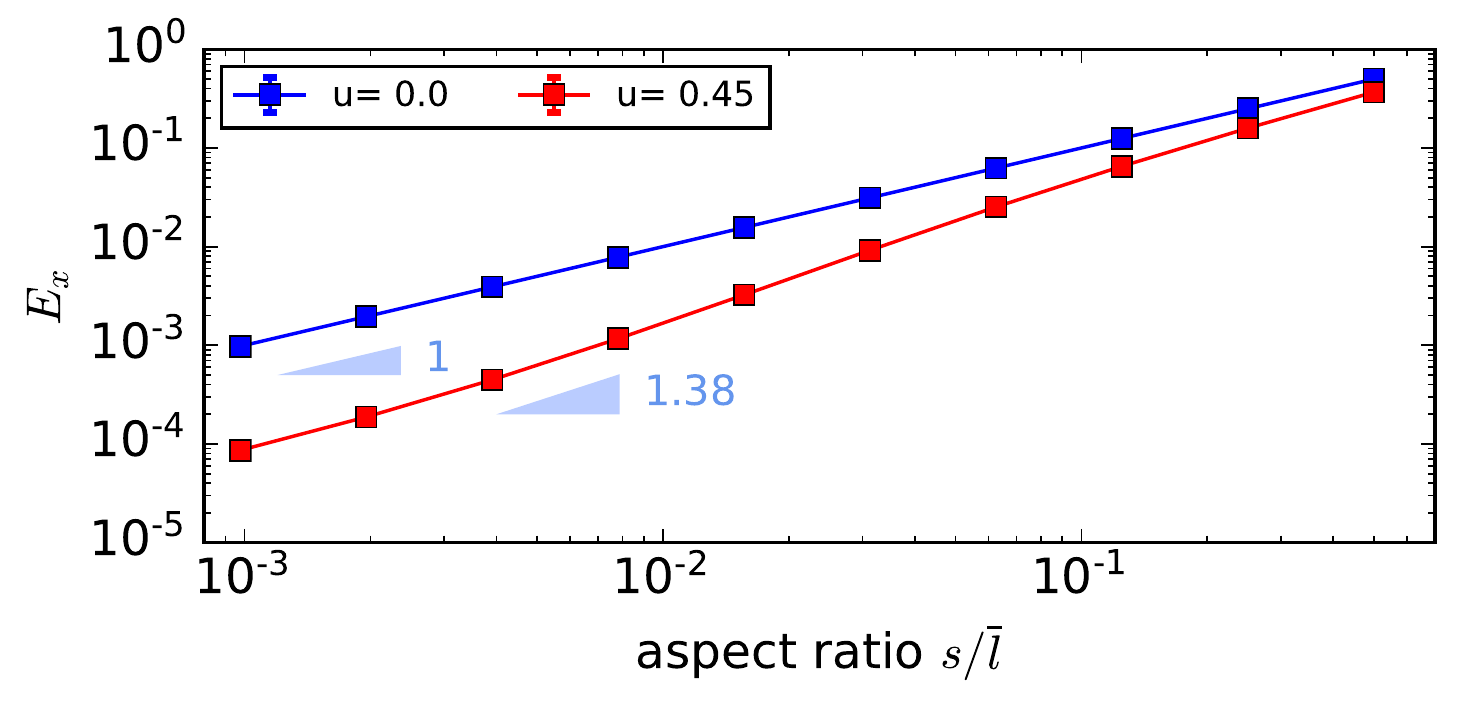}
\caption{(color online) Young's modulus along the x-axis $E_{x}$ as a function of beam aspect ratio $s/\ell_0$  for increasing disorder $u$ in the square-based lattices. The different curves correspond to different values $u$: $u=0$ for the blue (upper) curve  and $u=0.45$ for red (lower) curve. Here, the specimen size is $R=32\ell_0$. $E_{x}$ is expressed in $E_s$ units and $u$ is expressed in $\ell_0$ units.}
\label{fig:S11_vs_ratio}
\end{figure}

Now let us shift to understanding how the material lattice compliance depends on the beam aspect ratio and disorder. Figure \ref{fig:S44_vs_ratio} presents the shear modulus $G = E_s/\overline{S}_{44}$ as a function of the aspect ratio $s/\ell_0$ for increasing disorder levels in the three investigated geometries (honeycomb, square, and triangular). In absence of disorder, the elasticity constants can be determined theoretically (Table \ref{table:E_nu_theory} and Ref. \cite{Gibson1997}), and the values from the simulations match theoretical predictions. 

In honeycomb-based lattices, $G \sim (s/\ell_0)^3$ (that is  $G\sim \rho^3$) independent of the disorder level. This is expected, as the node connectivity ($Z=3$) does not satisfies Maxwell's condition for rigidity. Hence, bending-dominated deformations dominate and elastic modulus scale as $(s/\ell_0)^3$ \cite{Deshpande2001}.

Similar to honeycomb-based lattices, increasing disorder does not modify the scaling $G\sim s/\ell_0$ in triangular-based lattices. In these cases, $Z=6$, which ensures that the pin-jointed version of the lattice is fully rigid; hence $G\sim s/\ell_0$ \cite{Deshpande2001}. 

The behavior of square-based disordered lattices is surprising. As disorder is introduced, the coexistence of two distinct scaling regimes is observed: $G\sim s/\ell_0$ at small aspect ratios and $G\sim (s/\ell_0)^3$ at large ones. On the contrary, a similar transition is not observed for $E_{x} = E_s / \overline{S}_{11}$ whose scaling exponent increases from 1 to $1.38$ as disorder is introduced. But this change happens with a unique regime (Fig. \ref{fig:S11_vs_ratio}).

\begin{figure}[htp]
\centering 
\includegraphics[width=\columnwidth]{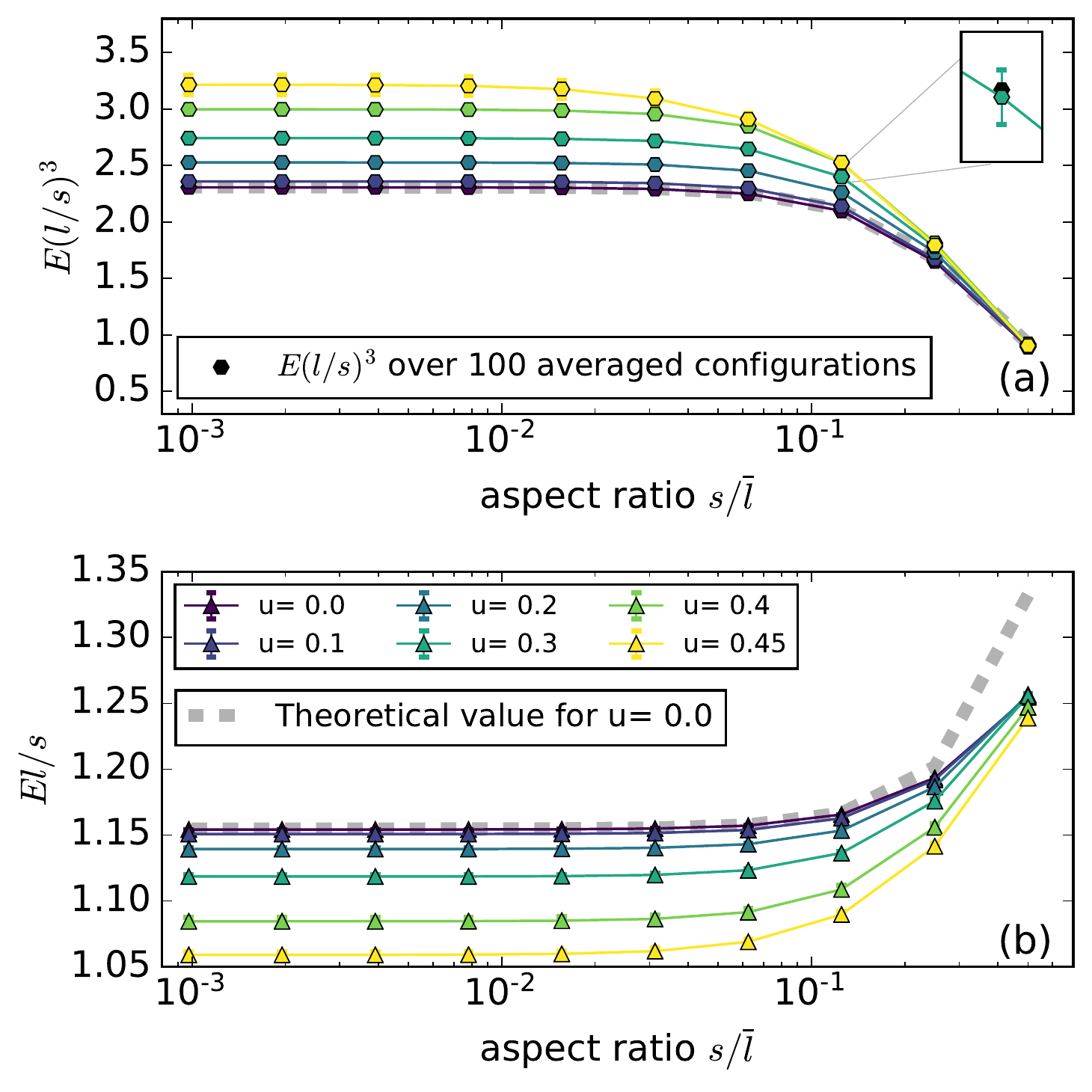}
\caption{(color online) Prefactor of scaling $E \sim (s/\ell_0)^n$ as a function of beam aspect ratio $s/\ell_0$  for increasing disorder $u$ in honeycomb-based lattices (panel a, $n=3$) and triangular-based lattices (panel b, $n=1$). The different colored curves correspond to different values $u$ according to the legend provided in panel c. Thick dash gray curve in panel a is the theoretical prediction $E/(s/\ell_0)^3 = (4\sqrt{3}/3) /(1+(5.4+1.5\nu_s)s^2/\ell_0^2)$ \cite{Silva1995}. Thick dash gray curve in panel b is the theoretical prediction $E/(s/\ell_0) = (2\sqrt{3})\times(1+s^2/\ell_0^2) /(3+s^2/\ell_0^2)$ \cite{Lipperman2007}. Here, specimen size is $R=40\ell_0$. $E$ is expressed in $E_s$ units and $u$ is expressed in $\ell_0$ units.}
\label{fig:E}
\end{figure}

All but the square-based lattices are mechanically isotropic. Hence, the elasticity behavior is fully characterized by two constants: Young's modulus and Poisson's ratio. Figure \ref{fig:E}(a) (resp. Figs. \ref{fig:E}(b)) presents $E/(s/\ell_0)^3$ (resp. $E/(s/\ell_0)$) versus $s/\ell_0$ for increasing architecture disorder in honeycomb-based lattices (resp. in triangular-based lattices). In absence of disorder, the curves coincide with the theoretically predicted value (Tab. \ref{table:E_nu_theory}) in the limit of slender beams (i.e. $s \ll \ell_0$). The plateau departure observed at larger $s/\ell_0$ is also fully consistent with the theoretical corrections provided in \cite{Silva1995,Lipperman2007} for thicker beams. In bending-dominated lattice (honeycomb-based lattice), increasing disorder stiffens the material and increases $E/(s/\ell_0)^3$ by $\sim 40\%$ for the maximum disorder ($u=0.45\ell_0$), see Fig. \ref{fig:E}(a). This variation is qualitatively consistent but quantitatively more pronounced than what is reported in the literature \cite{Mukhopadhyay2017, Silva1995,Fazekas2002,Li2005}. The root cause of these differences lies with how disordering occurs, by displacing randomly the points, the Voronoi tessellation of which provides the initial periodic honeycomb lattice. In stretching-dominated lattices (triangular-based lattices), increasing disorder yields softening effect and  decreases $E/(s/\ell_0)$ by respectively $\sim 8\%$ and $\sim 12\%$ for maximum disorder (Fig. \ref{fig:E}(b)). This is consistent with the finite element observations reported in \cite{Symons2008} on imperfect triangular lattices.

\begin{figure}[htp]
\centering 
\includegraphics[width=\columnwidth]{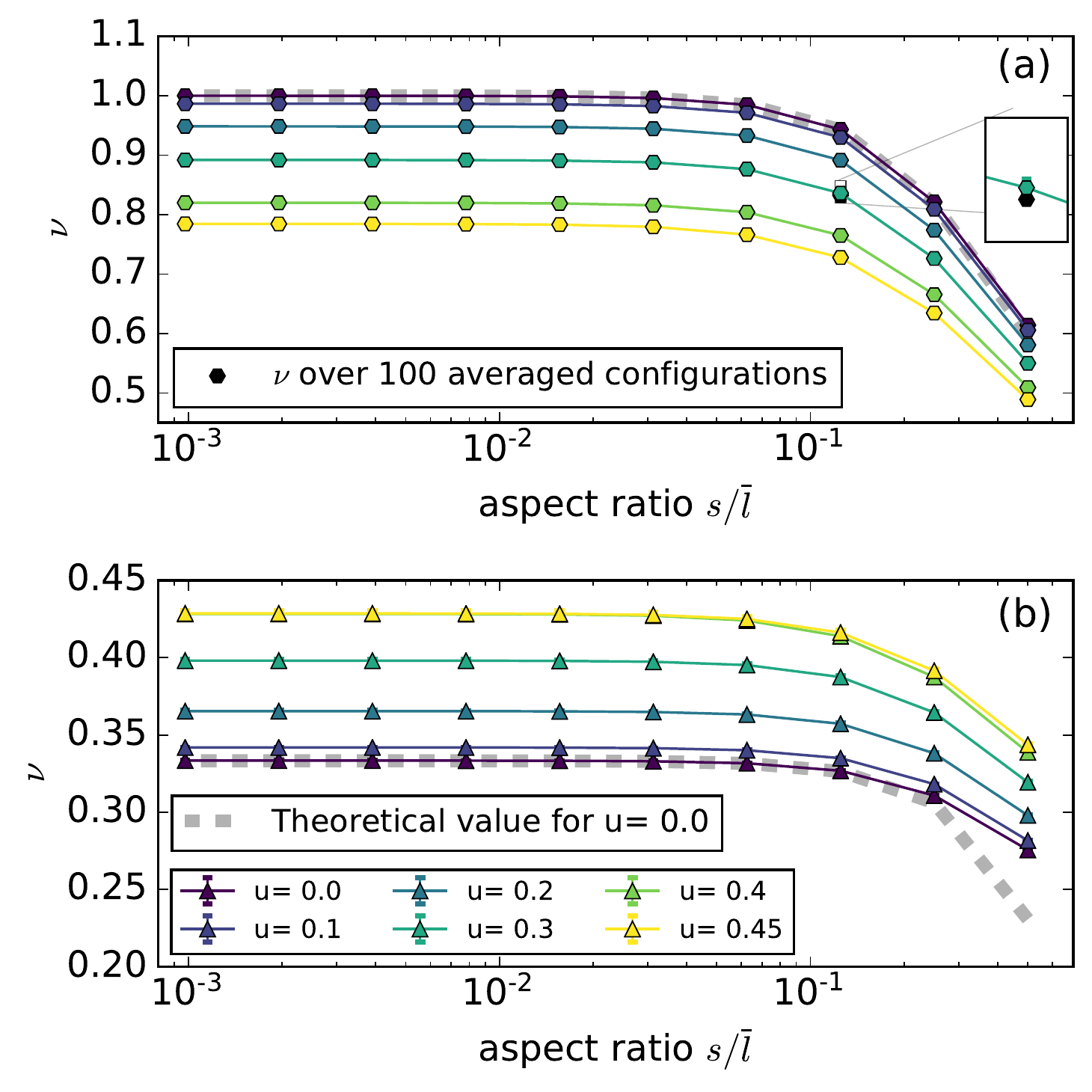}
\caption{
(color online) Poisson's ratio $\nu$ as a function of beam aspect ratio $s/\ell_0$  at increasing disorder $u$ in honeycomb-based lattices (panel a) and triangular-based lattices (panel b). The different colored curves correspond to different values $u$ according to the legend provided in panel b. 
Thick dash gray curve in panel a is the theoretical prediction $\nu = (1+(1.4+1.5\nu_s)s^2/\ell_0^2)/(1+(5.4+1.5\nu_s)s^2/\ell_0^2)$ \cite{Silva1995}. Thick dash gray curve in panel b is the theoretical prediction $\nu = (1/3)\times(1-s^2/\ell_0^2) /(1+s^2/3\ell_0^2)$ \cite{Lipperman2007}. 
Here, specimen size is $R=40\ell_0$. $u$ is expressed in $\ell_0$ units.}
\label{fig:nu}
\end{figure}

Figure \ref{fig:nu} shows the variations of Poisson's ratio, $\nu$, as a function of $s/\ell_0$ for the honeycomb- (panel (a)) and triangular-based (panel (b)) lattices. Like $E$ versus $s/\ell_0$ curves (and probably for the same reason), plateaus occur for $s/\ell_0 \ll 0.1$, and a departure exists for higher values. In honeycomb-based lattices, $\nu$ starts from the theoretically predicted value $\nu=1$ in absence of disorder, and decreases to $\nu \simeq 0.78$ as disorder increases to $u=0.45$. This decrease is significantly larger than that reported in Refs. \cite{Silva1995,Fazekas2002,Li2005}. In triangular-based lattices, $\nu$ starts at the theoretically predicted value $\nu=1/3$ and increases up to $\nu \simeq 0.42$ as disorder level increases to $u=0.45$.

\section{Discussion and analytical analysis}\label{Sec4}

This numerical study was designed to shed light on how the introduction of disorder in the architecture of lattice materials modifies their elasticity response. In this context, we derived and validated a novel procedure to determine the spatial distribution of Hooke's softness tensor at the local (joint) scale in 2D beam networks of prescribed architecture. Introducing disorder yields large spatial variations for local elasticity constants. Nevertheless, the correlation length associated with these spatial variations is small, approximately the beam length. Averaging them over the specimen provides an accurate estimation of the continuum-level scale values. 

A first effect of disordering is to promote elasticity isotropy when the parent crystalline architecture exhibits anisotropic elasticity (Fig. \ref{fig:UAI_vs_disorder}). This is expected since increasing disorder attenuates the rotation axis inherent to the pristine crystalline lattice and makes it more and more statistically invariant upon rotation.

Beyond isotropy, introducing disorder modifies the continuum-level (global) scale elastic constants in a way that depends deeply on the parent periodic geometry and connectivity. In summary:  

\begin{itemize}
\item[I] In lattices of high and low connectivity like triangular-based and honeycomb-based ones respectively, increasing disorder does not modify the scaling between elastic modulus and $s/\ell_0$ (or equivalently $\rho$): $E \sim s/\ell_0$ in highly connected lattices ($Z=6$) and $E \sim (s/\ell_0)^3$ in weakly connected ones ($Z=3$), no matter how much disorder is introduced.
\item[II] Increasing disorder softens highly connected (triangular) lattices; the prefactor, $E/(s/\ell_0)$ decreases with the amount of disorder, $u$.   
\item[III] Increasing disorder stiffens weakly connected (honeycomb) lattices, with a prefactor, $E/(s/\ell_0)^3$ decreasing with $u$.
\item[IV] Introducing disorder in lattices of intermediate connectivity like square-based ones ($Z=4$) dramatically modifies the scaling between shear modulus and $s/\ell_0$. Indeed, in the absence of any disorder, $G \sim (s/\ell_0)^3$ over the whole range, but a novel scaling regime $G \sim s/\ell_0$ ocurs at small $s/\ell_0$ values as soon as disorder is introduced.  
\end{itemize}

\subsection*{On the effect of disorder on $E\,vs. \,s/\ell$ scaling at low and high connectivity}

As already mentioned in Sec. III.D, observation I is explained by the fact that introducing disorder, here, does not change lattice connectivity. In disordered triangular-based lattices, the connectivity is always sufficient ($Z\geq 6$) to prevent collapsing mechanisms in the pin-jointed version; hence deformations, are always stretching-dominated \cite{Deshpande2001}. Similarly, in honeycomb-based lattices, $Z=3$ no matter how much disorder is introduced; Maxwell's criterion for stability is not fulfilled and, as such the structure remains bending-dominated at any value $u$.

\subsection{Local vs global elasticity constants in disordered lattices}
\label{sec:locvsglob}

\begin{figure}[!t]
\centering 
\includegraphics[width=\columnwidth]{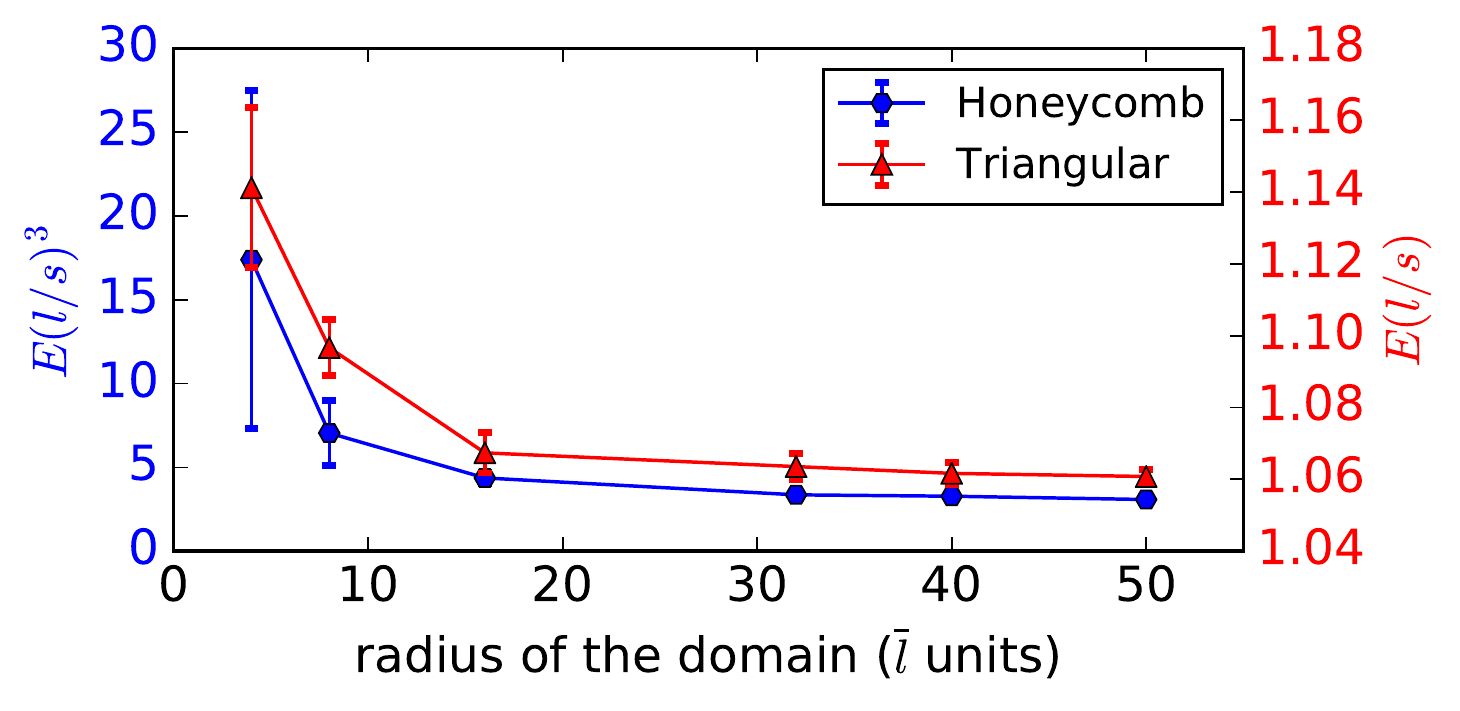}
\caption{(color online) Prefactor of Young's modulus for hexagonal and triangular lattices as function of the size of the sample. Lattices are disordered at $u=0.45$ and have an aspect ratio of $s / \ell_0 = 1/1024$.}
\label{fig:convergence}
\end{figure}

Observations II and III are more counter-intuitive. In particular, observation II is {\em opposite} to what would have been analytically predicted  for {\em local} elastic modulus, at the cell scale, after averaging over disorder configurations (Appendix B). This highlights that, as soon as disordered lattices are considered and irrespectively of the amount of disorder (even for arbitrary small ones), spatial correlation in stress redistribution should be taken into account and a large enough specimen should be defined. Figure \ref{fig:convergence} shows how the prefactor $E/(s/\ell_0)$ (resp. $E/(s/\ell_0)^3$) evolves with the specimen radius $R$ in disordered triangular-based (resp. honeycomb-based) lattices. Specimens of radius $R > 30 \ell_0$ should be prescribed to ensure that the determined elasticity moduli are truly material constants, independent of $R$.   

\begin{figure}[!t]
\centering 
\includegraphics[width=\columnwidth]{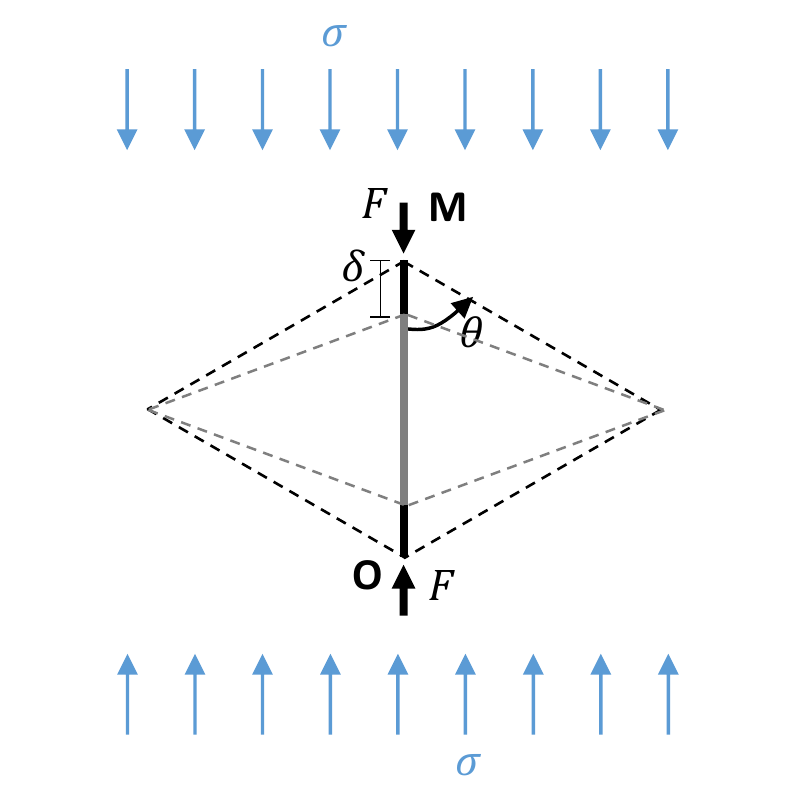}
\caption{(color online) (a) Sketch of beam deformation in a streching-dominated lattice loaded under a compressive stress $\sigma$. The stretching force applying onto the considered beam is $F = \sigma \ell s \mathcal{F}(\theta)$ where $\theta$ is the typical angle between two joint beams and $\ell \mathcal{F}(\theta)$ is the typical distance separated two successive vertically stretched beams. In response to this stretching force, each edge of the beam moves over a distance $\delta$. Thick black vertical plain line and dotted black inclined lines present the considered beam and two jointed ones before the lattice deformation while gray vertical thick plain line and gray dotted inclined lines show the same beams after deformation.}
\label{fig:sketch_stretching}
\end{figure}

\subsection{On the origin of disorder-induced softening in stretching-dominated lattices}

Because of the nonlocality mentioned in the previous section, observation II is extremely difficult to rationalize quantitatively. It has been shown in Ref. \cite{Gurtner2014}, that the stiffness of stretching-dominated lattices decreases when geometrical disorder yields non-affine strain fields. Here, we propose an alternative explanation and argue that the observed softening is due to the increase of mean beam length as disorder increases. 

To some extent, this can be rationalized by considering a given beam,  $\OO\M$, and a constant compressive stress $\sigma$ applying parallel to it  $\OO\M$ (Fig. \ref{fig:sketch_stretching}). The force $F$ applying at each edge of the beam is $F = \sigma \ell s \mathcal{F}(\theta)$ where $\ell$ is the beam length, $\theta$ is the typical angle between two joint beams (considered to be the same everywhere, e.g. $\theta=\pi/6$ in disorder-free triangular lattice), and $\ell \mathcal{F}(\theta)$ is the typical distance separating two successive beams compressed by $\sigma$ (Fig. \ref{fig:sketch_stretching}). Each node, $\OO$ and $\M$, moves by $\delta= F \ell/ E_s s^2 = \sigma \ell^2 \mathcal{F}(\theta)$, due to beam contraction. This yields a strain $\epsilon = \delta / \ell = \sigma \ell \mathcal{F}(\theta)/E_s s$. Finally, Young's modulus $E^{(str)} = \sigma/\epsilon$ is given by:

\begin{equation}
	E^{(str)} = \frac{E_s s}{\ell \mathcal{F}(\theta)}
	\label{Eq:Sec4-2}
\end{equation}

Now, disorder is introduced on  edge position:  $\xx(\M) = \xx_0(\M) + u\mathbf{\eta} (\M)$, where ${\bf \eta}(\M)$ is given by:

\begin{equation}
\eta(\M) = \x (\cos \theta_{\M} - \cos \theta_{\OO})+ \y (\sin \theta_{\M}-\theta_{\OO}),
	\label{Eq:Sec4-3}
\end{equation}

\noindent where $\theta_{\M}$ and $\theta_{\OO}$ are two angles selected randomly between $-\pi$ and $\pi$. This yields fluctuations of the beam length, which, to second order terms in $u$, now writes $\ell = \ell_0+u \eta_y(\M) +u^2\eta^2_x(\M)/(2\ell_0) +o(u^2/\ell^2_0) $. As a result, the strain also fluctuates in space. By averaging $\epsilon$ over different configurations $\{\theta_{\OO},\theta_{M}\}$, one gets $\overline{\epsilon} = \sigma\overline{\ell} \mathcal{F}(\theta)/E_s s$, and the effective Young's modulus becomes:

\begin{equation}
	\overline{E}^{(str)} = \frac{E_s s}{\overline{\ell} \mathcal{F}(\theta)}
\end{equation}

\noindent To the second order in $u$, the  mean beam length is $\overline{\ell}(u) = \ell_0+1/2 u^2/\ell_0+o(u^2/\ell^2_0)$, and finally:

\begin{equation}
	\overline{E}^{(str)}(u) = E^{(str)}_0\left( 1 - \frac{1}{2}\frac{u^2}{\ell_0^2} + o\left(\frac{u^2}{\ell^2_0}\right)\right)
	\label{Eq:Sec4-6}
	\end{equation}

\begin{figure}[tpb]
\centering 
\includegraphics[width=\columnwidth]{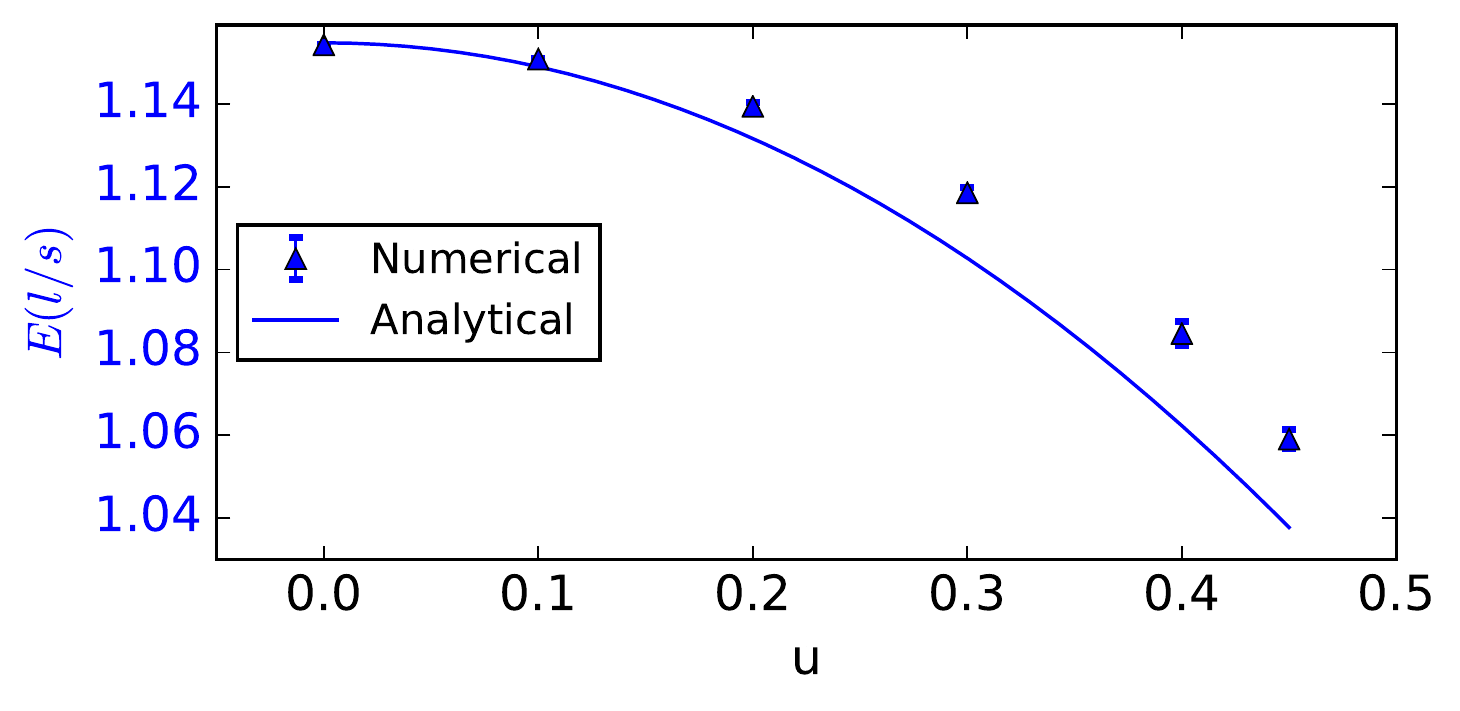}
\caption{(color online) Young's modulus prefactors for triangular lattices computed numerically (black triangles) and estimated through analytical second order's expansion (Eq. \ref{Eq:Sec4-6}), with $E^{(str)}_0 = 2 \sqrt{3}s/\ell_0$ (Tab. II).}
\label{fig:analytical_tri_hx}
\end{figure}

\noindent As shown in Fig. \ref{fig:analytical_tri_hx}, this analytical estimate is consistent with numerical observations in triangular-based disordered lattices, as least as long as $u$ is not too large.

\subsection{On the origin of disorder-induced stiffening in bending-dominated lattices}

{\em A priori}, the same argument can be applied to bending-dominated lattices. As $E^{bnd} \sim (s/\ell)^3$, this would yield $\overline{E}^{bnd}(u) =  E_0^{bnd}( 1 - 3/2 u^2/\ell^2_0 + o(u^2/\ell^2_0)$; hence, the lattice would have been expected to soften as disorder increases. Observation III is the opposite.  

We argue that the disorder-induced stiffening observed in bending-dominated lattices translates the fact that, in any bending-dominated lattice, part of applied stress is accommodated by beam stretching. Indeed, as force applied on a given node, its projection parallel to the considered beam yields beam dilation or stretching while the part perpendicular to this beam yields bending. As disorder increases, the relative importance of the stretching part increases. Assuming that the lattice deformation due to strecthing are negligible with respect to those due to bending, the total lattice deformation is fully governed by the part of applied stress accomodated by beam bending, only. Then, as the Young's modulus $E^{(bnd)}$ is set by the total force over the total deformation ratio, $E^{(bnd)}$ is set by the total force over effective bending force ratio, which increases as disorder increases. 

\subsection{On the two elasticity scaling regimes in disordered square lattices}

\begin{figure}[!t]
\centering 
\includegraphics[width=\columnwidth]{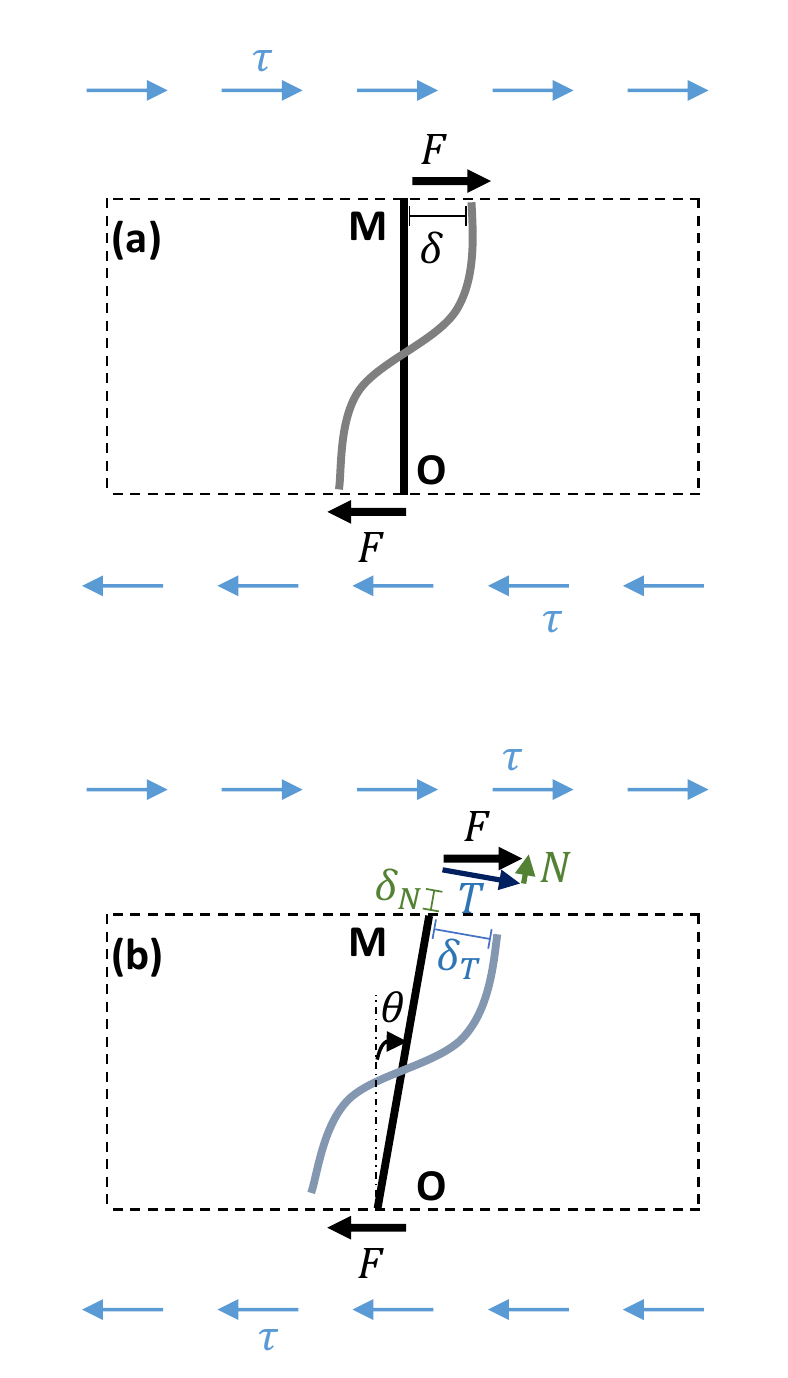}
\caption{(color online) (a) Sketch of beam deformation in a square lattice loaded under a shear stress $\tau$. Due to the shear, the force applying onto the considered beam is $F = \tau \ell$. In response to $F$, the beam bends and its two edges moves over a distance $\delta$. Thick black vertical plain line and dotted black inclined lines present the considered beam and two jointed one before the lattice deformation while gray vertical thick plain line and gray dotted inclined lines show the same beams after deformation. (b) Due to disorder, the considered beam makes now an angle $\theta$ with respect to the perpendicular of shear. Applied force $F$ then splits into a component $T$ perpendicular to the beam and a parallel one, $N$. The former makes the beam bend and its two edges move over a distance $\delta_T$ and the former makes the beam stretch and its two edges move over a distance $\delta_N$.}
\label{fig:sketch_square}
\end{figure}

As in honeycomb-based bending-dominated lattices, the appearance of two scaling regimes in the $G\, vs. \, s/\ell$ curves (Fig. \ref{fig:S44_vs_ratio}b) in disordered square lattice is also attributed to the fact that an increasing proportion of applied shear stress is accommodated via latttice stretching as introduced disorder increases. In this specific case, this can be rationalized as follow. 

Let us first consider the case of disorder-free square lattice loaded by a constant shear stress, $\tau$, applied perpendicularly to a given beam $\OO \M$ (Fig. \ref{fig:sketch_square}). Force $F$ exerted on node $\M$ is $F = \tau s \ell$, and it applies perpendicular to the beam. Due do beam bending, $\M$ moves over a distance $\delta = F \ell^3 /( E_s s^4)  = \tau \ell^4 / (E_s s^3)$. $\OO$ moves along the opposite direction over the same distance, and the resulting elementary shear deformation $\gamma = 2\delta/\ell = \tau 2 \ell ^3 /E_s s^3$ is:

\begin{equation}
\gamma = \frac{2 \ell^3}{E_s s^3} \tau
\label{Eq:Sec4-7}
\end{equation}

Let us now introduce disorder on edge positions. This yields fluctuations on the angle $\theta$ between $\OO \M$ and the vertical to applied shear (Fig. \ref{fig:sketch_square}b). As a result $F$, is not perpendicular to $\OO \M$ anymore. The perpendicular component $T = F \cos\theta$ makes the beam bend and $\M$ moves over a distance $\delta_T = F\cos\theta \ell^3 /( E_s s^4)$ along a direction perpendicular to $\OO \M$; the parallel component $N = F \sin \theta$ makes the beam stretch and $\M$ moves over a distance $\delta_N = F\sin\theta \ell/(E_s s)$ in a direction parallel to $\OO \M$. The former implies a bending-induced elementary shear deformation $\gamma^{(bnd)} = \delta_T\cos\theta/\ell = 1/2\tau \cos^2\theta \ell^3/(E_s s^3)$; the latter implies a shear-induced elementary shear deformation $\gamma^{(str)} = \delta_N\sin\theta/\ell = 1/2\tau \sin^2\theta \ell/(E_s s)$. 

Let us now assume that disorder is introduced by moving lattice edge over a random displacement as stipulated in Eq. \ref{Eq:Sec4-3}. To the second order in $u/\ell_0$,  $\cos^2\theta = 1- 1/2 \eta^2_x u^2/\ell^2_0 + o(u^2/\ell^2_0)$ and $\sin^2\theta = 1/2 \eta^2_x u^2/\ell^2_0 + o(u^2/\ell^2_0)$. By averaging over the configurations $\{\theta_\OO,\theta_\M\}$, one gets:

 \begin{equation}
 \begin{split}
\overline{\gamma}^{(bnd)}(u) & = \tau \frac{2}{E_s}\frac{\ell^3_0}{s^3} 
\left(1-\frac{1}{2}\frac{u^2}{\ell^2_0}+ o\left(\frac{u^2}{\ell^2_0}\right)\right)\\
\\
\overline{\gamma}^{(str)}(u) & = \tau \frac{2}{E_s}\frac{\ell_0}{s} \left(\frac{1}{2}\frac{u^2}{\ell^2_0} + o\left(\frac{u^2}{\ell^2_0}\right)\right) 
 \end{split}
\end{equation}

\noindent Finally, the averaged shear-induced and bending-induced elementary shear moduli $\overline{G}^{(str)}=\tau/\overline{\gamma}^{(str)}$ and $\overline{G}^{(bnd)}=\tau/\overline{\gamma}^{(bnd)}$ become:

 \begin{equation} 
 \begin{split}
 \overline{G}^{(bnd)}(u) & = \frac{E_s}{2}\frac{s^3}{\ell^3_0} \left(1+\frac{1}{2}\frac{u^2}{\ell^2_0}+ o\left(\frac{u^2}{\ell^2_0}\right)\right)\\
 \overline{G}^{(str)}(u) & = E_s \frac{s}{\ell_0} \left(\frac{u^2}{\ell^2_0}+ o\left(\frac{u^2}{\ell^2_0}\right)\right)
\label{Eq:Sec4-8}
\end{split}
\end{equation}

\begin{figure}[!t]
\centering 
\includegraphics[width=\columnwidth]{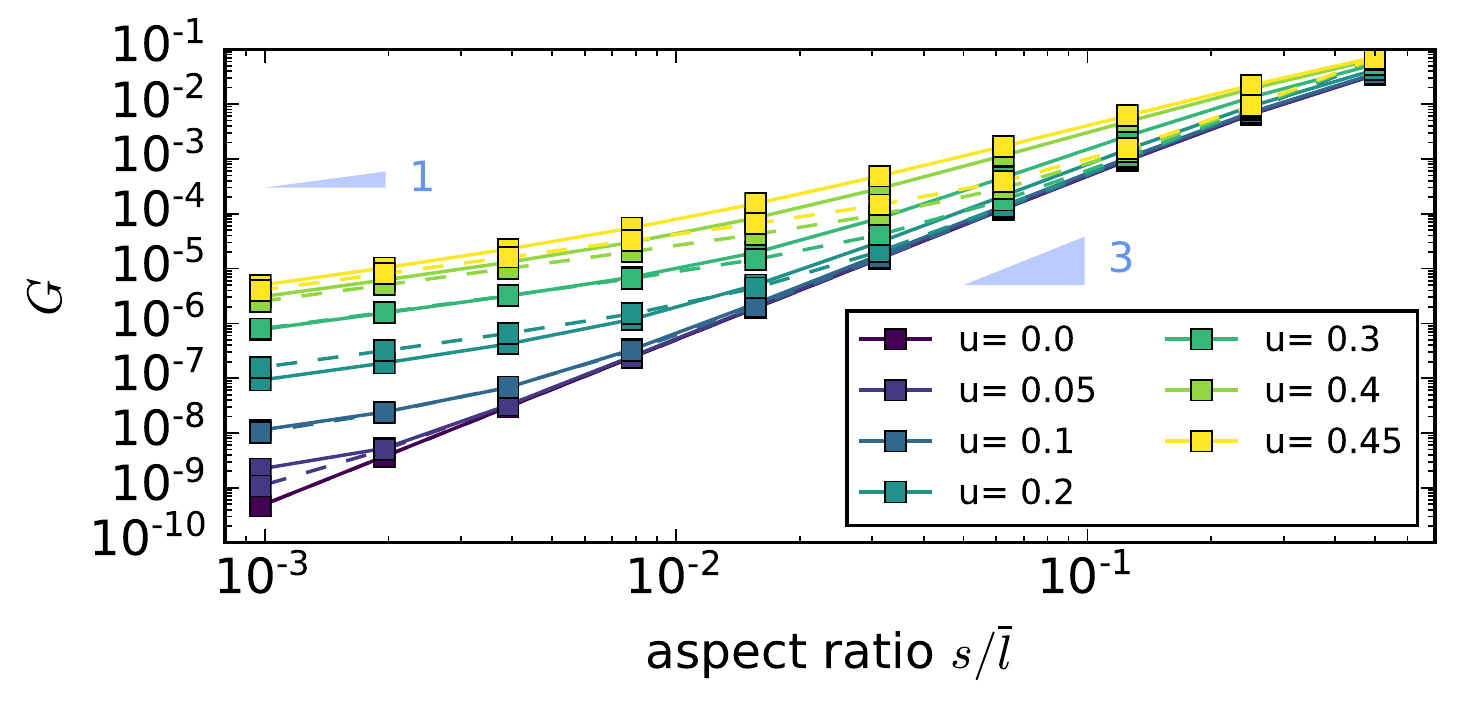}
\caption{(color online) Shear modulus $G$ as a function of beam aspect ratio $s/\overline{\ell}$ at increasing disorder $u$ in square-based disordered lattices. Colored symbols connected by plain lines are the curves obtained via simulation. Colored dash lines are the analytical expression given by Eq. \ref{Eq:Sec4-9} with $\alpha=0.1$. $G$ is expressed in $E_s$ units and $u$ is expressed in $\overline{\ell}$ units.}
\label{fig:G_sq_analytical}
\end{figure}

The difficulty is to go from these local, configuration-averaged, shear moduli to the global specimen-scale shear moduli, $G^{(sq)}$. As discussed in Sec. \ref{sec:locvsglob}, this upscaling is not trivial and cannot be obtained quantitatively via the simple addition of the two contributions. 
Still, making $G^{(sq)} = \overline{G}^{(bnd)} + \overline{G}^{(str)}$ allows reproducing at least qualitatively the observed features: When $s/\ell \ll u/\ell$, $ \overline{G}^{(str)}(u) \gg  \overline{G}^{(bnd)}(u)$ and $ G^{(sq)}(u) \approx \overline{G}^{(str)}(u) \sim s/\ell$; conversely, when $s / \ell \gg u/\ell$, $\overline{G}^{(str)}(u) \ll  \overline{G}^{(bnd)}(u)$ and $ G^{(sq)}(u) \approx \overline{G}^{(bnd)}(u) \sim s^3/\ell^3$. 
This qualitatively explain observation IV and Fig. \ref{fig:S44_vs_ratio}b, with a linear scaling regime $G \sim s/\ell$ at small aspect ratios, a cube scaling regime $G\sim s^3/\ell^3$ at large aspect ratios, and a crossover aspect ratio increasing as disorder increases. Actually, as shown on Fig. \ref{fig:G_sq_analytical}, all the numerical curves obtained in disordered square lattices are reproduced quantitatively using a weighted sum:

 \begin{equation} 
 \overline{G}^{(sq)}(u)  = (1-\alpha u^2) \overline{G}^{(bnd)}(u) + \alpha u^2 \overline{G}^{(str)}(u)
\label{Eq:Sec4-9}
\end{equation}

\noindent with $\alpha \simeq 0.1$.

\section{Conclusion}\label{Sec5}

The series of simulations reported here investigated how the introduction of disorder modifies the elasticity behavior of 2D lattice materials of various architectures. A procedure inspired from the modeling of granular systems has been developed to determine the map of the full elasticity tensor at the local scale. This procedure has been validated via comparisons with theoretical results known for pure crystalline lattices. Introducing disorder has the disadvantage of generating important spatial fluctuations on these elastic constants. Nevertheless, the associated correlation length remains small, on the order of the beam length $\ell$. Averaging over the specimen provides an accurate determination of the continuum-level scale elasticity constants (compliance tensor).   

First, as demonstrated here on 2D square-based lattice materials, introducing disorder in a crystalline architecture of initial anisotropic elasticity promotes elasticity isotropy. Note that, while there are elastically isotropic crystalline architectures in 2D (e.g. the triangular or honeycomb lattice studied here), this is no longer the case in 3D \cite{Ranganathan2008}. Lattice structures with isotropic elasticity are important for many applications and, as such, are the subject of several works \cite{Gurtner2014,Xu2016,Berger2017,Tancogne-Dejean2018,Latture2018}. Introducing disorder in a tunable way as proposed here may offer a promising route to this aspect.  

Second, introducing disorder softens highly-connected stretching-dominated lattice materials. This observation is somehow counter-intuitive since a simple argument, based on the analysis of the elastic energy stored in an elementary cell of the parent crystalline lattice and the evolution of this configuration-averaged energy in presence of disorder would have predicted the opposite (appendix B). This highlights the importance of nonlocality (and the induced difficulty to anticipate the effect of disorder) on the elasticity properties of lattice materials. This also relates to previous studies \cite{Gurtner2014} that evidence correlations between network stiffness and disorder-induced non-affine strain fields in stretching-dominated lattices.

Third, introducing disorder helps stiffening low connectivity bending-dominated lattice materials. The effect is quite small in honeycomb lattices: elastic modulus varies as the cube of density (driven by changing the beam aspect ratio) regardless of the disorder level, and disorder only plays on the prefactor, which increases by $\sim 40\%$. On the other hand, the effect is drastic in square-based lattices. In this scenario, stiff regime (linear scaling between shear modulus and density) is observed at low density, whilst the soft regime observed in absence of disorder (cubic scaling between shear modulus and density) is only recovered at large density. Additionally, the crossover density between these two regimes is selected by the disorder level. We conjecture that similar features will be observed in any lattices where connectivity is too low to ensure full structure rigidity ($Z < 6$ in 2D, $Z < 12$ in 3D), but large enough to get rigid local cells ($Z > 3$ in 2D, $Z > 4$ in 3D). 

On-going work aims at assessing this conjecture. If this were the case, modulating spatially the disorder introduced in an initially crystalline architecture would provide a promising way to obtain meta-composites made of soft and stiff zones, the spatial entanglement of which could be arranged. This may allow the design of novel architectures for materials with both large stiffness and energy-absorbers, which are a priory antagonist.

\section*{Acknowledgments}

Support through the CEA PTC Materiaux \& Procédés (project LightToughMetaMat) is gratefully acknowledged. 

\appendix

\section{On the effect of averaging procedure onto macroscale compliance tensor}
\label{A1}

Compliance tensor can be calculated following two schemes. First, $\overline{\MS}$ is classically obtained using macroscopic stress and strain. Second, a local $\MS_N(x,y)$ is computed by first coarse-graining local stress and strain fields over square meta-cell of edge size $N\ell_0$ which gives $\stress_N(x,y)$ and $\strain_N(x,y)$ and then using the classical best-fit procedure.  

\begin{figure}[!t]
\centering 
\includegraphics[width=\columnwidth]{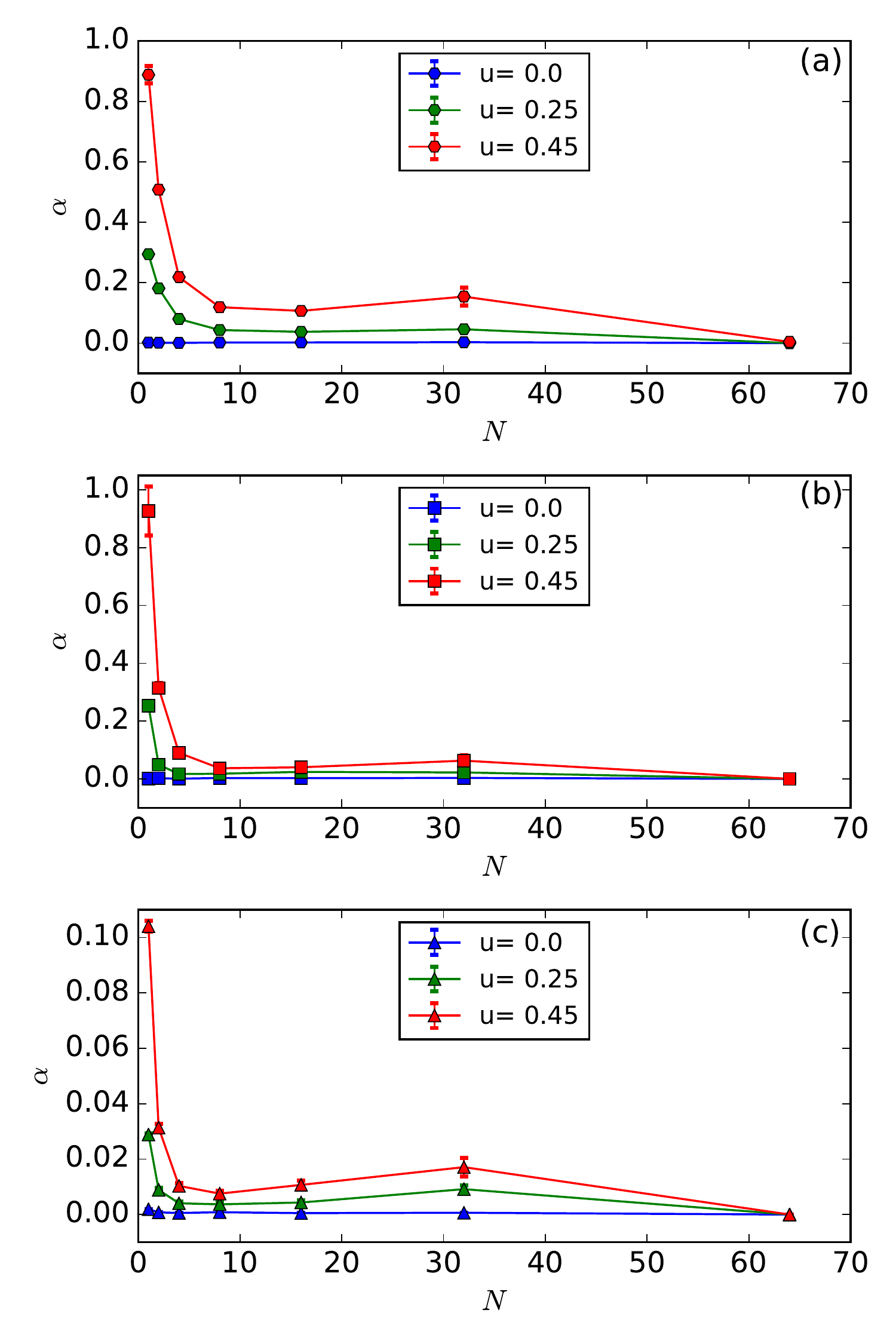}
\caption{(color online) Ratio $\alpha$ (Eq. \ref{eq:alpha}) as a function of coarse-graining scale $N$, at increasing levels of disorder $u$, in  honeycomb-based lattices (panel a), square-based lattices (panel b) and triangular-based lattices (panel c). In each panel, the three curves correspond to $u=0$ for the blue (lower) curve, $u=0.25$ for the green (intermediate) curve and $u=0.45$ for the red (upper) curve. Here, beam aspect ratio is $s/\ell_0=1/8$ and specimen size is $R=32\ell_0$. $u$ is expressed in $\ell_0$ units.}
\label{fig:VER_vs_N}
\end{figure}

A coefficient $\alpha$ is defined as: 

\begin{equation} 
\alpha = \frac{\left\| \overline{\MS} - \langle \MS_{N} \rangle \right\|}{\left\|\overline{\MS} \right\|}
\label{eq:alpha}
\end{equation} 

\noindent where $\langle \rangle$ denotes the specimen average and $||\hspace{0.05cm}||$ the Euclidean norm. Figure \ref{fig:VER_vs_N} shows the variation of $\alpha$ with $N$ in the different geometries studied here. Like $g(r)$, $\alpha(r)$ decreases rapidly to zero. Hence, it is equivalent to average over space or over configurations for compliance tensor components.

\section{Analytical expression of local configuration-averaged elastic moduli in disordered triangular-based lattices}
\label{A2}

The idea is based on the equivalence between the continuum-level scale elastic strain energy stored in a unit cell, $U_{continuum}$ and the total energy stored in the beams of this cell, $U_{cell}$ \cite{Chen1998,Ostoja-Starzewski2002}. $U_{continuum}$ is given by:

 \begin{equation}
U_{continuum} = \frac{1}{2} \epsilon_{ij} C_{ijkl} \epsilon_{kl},
\label{Eq:Ucontinuum}
 \end{equation}
 
\noindent Where $C_{ijkl}$ are the components of stiffness tensor and $\epsilon_{ij}$ are the components of strain tensor. Here and in the following, Einstein summation convention is employed on repeated indices. Moreover, $\epsilon_{ij}$ are considered to be uniform throughout the lattice. $U_{cell}$ is given by:

 \begin{equation}
U_{cell} = \frac{1}{4 S_{vor} s}\sum_{p}^{Z} F_i(p) U_i(p),
\label{Eq:Ucell}
 \end{equation}
 
\noindent where $S_{vor}$ is the area of the Voronoï polyhedron around the considered node (labeled $\OO$), $p$ index runs over all beams starting from this node, $\F(p) = \F_{\M_p \rightarrow \OO}$, and $\U(p) = \U(\M_p) - \U(\OO)$ is the relative displacement of node $\M_p$ with respect to node $\OO$. Note the factor $1/4$ (and not $1/2$) in Eq. \ref{Eq:Ucell} that comes from the fact that the energy of each beam equally splits in the two Voronoi cell associated to each edge. as strain $\epsilon_{ij}$ are uniform:

\begin{equation}
U_i(p) = \epsilon_{ij} x_j(p),
\label{Eq:UvsEps}
 \end{equation}

\noindent where $\xx(p) = \xx(\M_p) -\xx(\OO)$  is the relative position of node $\M_p$ with respect to node $\OO$. Due to their high connectivity, deformation in triangular-based lattices are dominated by beam stretching. Hence, $\F(p)$ is:
 
 \begin{equation}
F_i(p) = E_s s^2 \frac{x_i(p)x_j(p)}{\ell(p)^3}U_j(p),
 \end{equation}
 
 \noindent where $\ell(p)$ is the length of beam $p$. Finally, one gets:
 
 \begin{equation}
C_{ijkl} = \frac{E_s s}{2 S_{vor}}\sum_{p}^{Z} \frac{x_i(p) x_j(b) x_k(p) x_l(p)}{\ell(p)^3}
\label{Eq:Cstrectch}
 \end{equation}
 
Let us first consider a disorder-free triangular lattice. Then, all beams have same length $\ell_0$, $S_{vor} = \ell_0^2\sqrt{3}/2$, and the relative positions $\xx_0(p)$ of the six nodes $\M_p$ are: $\{\ell_0,0\}$, $\{\ell_0/2,\ell_0\sqrt{3}/2\}$, $\{-\ell_0/2,\ell_0\sqrt{3}/2\}$, $\{-\ell_0,0\}$, , $\{-\ell_0/2,-\ell_0\sqrt{3}/2)\}$ and  $\{\ell_0/2,-\ell_0\sqrt{3}/2\}$. As a result, we get:

 \begin{equation}
{\bf C}_0(tr)= \frac{3}{4\sqrt{3}} \frac{E_s s}{\ell_0}
\begin{bmatrix} 
3 & 1 & 0 \\
1 & 3 & 0 \\
0 & 0 & 1 \\
\end{bmatrix},
\end{equation}

\noindent This compliance matrix is that of a linear elastic isotropic material of Young's modulus and Poisson's ratio given by: 

 \begin{equation}
E_0(tr)= \frac{2}{\sqrt{3}} \frac{E_s s}{\ell_0}, \quad \nu_0(tr)= \frac{1}{3}
\end{equation}

Now, disorder is introduced on the positions:  $\x(p) = \x_0(p) + u{\bf \eta} (p)$ where ${\bf \eta}(p)$ is given by Eq. \ref{Eq:Sec4-3}. These changes are introduced in Eq. \ref{Eq:Cstrectch} and the compliance matrix is deduced:

   \begin{align}
C_{11} =  \frac{E_s s}{2 S_{vor}}\sum_{p}^{Z} \frac{(x_0(p)+u\eta_x(p))^4}{((x_0(p)+u\eta_x(p))^2+(y_0(p)+u\eta_y(p))^2)^{3/2}} \nonumber\\
C_{22} =  \frac{E_s s}{2 S_{vor}}\sum_{p}^{Z} \frac{(y_0(p)+u\eta_y(p))^4}{((x_0(p)+u\eta_x(p))^2+(y_0(p)+u\eta_y(p))^2)^{3/2}} \nonumber \\
C_{12} =  \frac{E_s s}{2 S_{vor}}\sum_{p}^{Z} \frac{(x_0(p)+u\eta_x(p))^2(y_0(p)+u\eta_y(p))^2}{((x_0(p)+u\eta_x(p))^2+(y_0(p))+u\eta_y(p))^2)^{3/2}} \nonumber\\  
C_{13} =  \frac{E_s s}{2 S_{vor}}\sum_{p}^{Z} \frac{(x_0(p)+u\eta_x(p))^3(y_0(p)+u\eta_y(p))}{((x_0(p)+u\eta_x(b))^2+(y_0(p)+u\eta_y(p))^2)^{3/2}} \nonumber\\ 
C_{23} =  \frac{E_s s}{2 S_{vor}}\sum_{p}^{Z} \frac{(x_0(p)+u\eta_x(p))(y_0(p)+u\eta_y(p))^3}{((x_0(p)+u\eta_x(p))^2+(y_0(p)+u \eta_y(p))^2)^{3/2}} \nonumber\\ 
 \end{align}

\noindent where Voight notation has been used and $S_{vor}$ has been assumed to remain unchanged by the disordering procedure. After expanding to second order in $u/\ell_0$ and subsequently averaging over disorder configurations so that $\overline{\eta_x} = \overline{\eta_y}= \overline{\eta_x \eta_y} = 0$ and $\overline{\eta^2_x} = \overline{\eta^2_y}  = 1$, we get:

\begin{equation}
{\bf C}(tr|u)= \frac{3}{4\sqrt{3}} \frac{E_s s}{\ell_0}
\begin{bmatrix} 
3 & 1 & 0 \\
1 & 3 & 0 \\
0 & 0 & 1 \\
\end{bmatrix}
\left(1 + \frac{u^2}{\ell^2_0} + o\left(\frac{u^2}{\ell^2_0}\right)\right)
\end{equation}

\noindent In the disordered triangular-based lattice, elasticity remains isotropic and local configuration-averaged Poisson's ratio at the elementary cell scale remains constant, $\nu(tr|u) = 1/3$. Conversely, local Young's modulus increases with disorder strength:

 \begin{equation}
E(tr|u)= E_0(tr)\left(1+\frac{u^2}{\ell^2_0}+ o\left(\frac{u^2}{\ell^2_0}\right)\right)
\end{equation}

\noindent This increase is opposite to the decrease observed for  global (specimen-scale) Young's modulus (Fig. \ref{fig:analytical_tri_hx}). Note that Eq. \ref{Eq:UvsEps} assume uniform strain at the local scale, which is not true in the presence of geometrical disorder \cite{Gurtner2014}.



%

\end{document}